\documentclass[twocolumn]{autart}    

\usepackage{graphicx}          
\usepackage{epstopdf}
\usepackage{amsmath} 
\usepackage{amssymb}  
\usepackage{listings}
\usepackage{algorithmic,algorithm}
\usepackage{color}

\usepackage{diagbox}
\usepackage{booktabs}

\newenvironment{example}{{\bf Example}\em}{\par}

\newcommand{\mR}{{\mathbb R}}

\newcommand{\mZ}{\mathbb Z}

\newcommand{\Frac}[2]{{\displaystyle\frac{#1}{#2}}}
\newcommand{\Hb}{\mathbf{H}}

\newcommand{\Zbb}{\mathbb Z}
\newcommand{\E}{{\mathbb E}}

\newcommand{\yb}{\mathbf{y}}

\newcommand{\rank}{{\rm rank}}
\newcommand{\Span}{{\rm span}}

\newcommand{\bmat}{\begin{bmatrix}}
\newcommand{\emat}{\end{bmatrix}}
\newcommand{\nd}{\noindent}

\begin{document}

\begin{frontmatter}

\title{Identification of Low Rank Vector Processes} 

\thanks[footnoteinfo]{An abridged version of this  paper,  \cite{PCL-21}, was   presented at  the  IFAC SYSID 2021 meeting in Padova.
}

\author[First]{Wenqi Cao}\ead{wenqicao@sjtu.edu.cn},
\author[Second]{Giorgio Picci}\ead{picci@dei.unipd.it},    
\author[Third]{Anders Lindquist}\ead{alq@math.kth.se}  

\address[First]{Department of Automation, Shanghai Jiao Tong University, Shanghai, China.}
\address[Second]{Department of Information Engineering, University of Padova, Italy.}  
\address[Third]{Department of Automation and School of Mathematical Sciences, Shanghai Jiao Tong University, Shanghai, China.}        


\begin{abstract}                          
We study modeling and identification of stationary processes with a spectral density matrix of  low rank. Equivalently, we consider processes having an innovation of reduced dimension for which Prediction Error Methods (PEM) algorithms are not directly applicable. We show that these processes admit a special feedback structure with a deterministic feedback channel which  can be used to split the identification in two steps, one of which can be  based on standard algorithms while the other is based on a deterministic least squares fit. Identifiability of the feedback system  is  analyzed and a unique identifiable structure is characterized. Simulations show that the proposed procedure works well in some simple examples.
\end{abstract}

\end{frontmatter}

\section{Introduction}\label{Introd}
Quite often in the  identification   of large-scale time series one has to deal with  {\em low rank} signals which   have a rank deficient spectral density. Such low rank time series may arise in diverse areas such as control systems, economics, networked systems, biology  and other fields.\\
Suppose we want to identify  an $(m+p)$-dimensional vector time series $y$  by modeling it as a weakly stationary zero-mean purely non deterministic (p.n.d.)  process  $y\equiv \{y(t)\,;\,t\in \Zbb\}$,   having  a rank deficient rational  spectral density $\Phi(z)$ of rank $m$.  This  spectral density  can always be written in factorized form
\begin{equation}\label{sfd}
  \Phi(e^{i\theta}) = W(e^{i\theta})W(e^{-i\theta})^{\top},
\end{equation}
with $W$  an  $(m+p)\times m$ full rank stable rational spectral factor. It is well-known that there are in general many such factors only one of which has the property of being {\em minimum phase}, see the appendix \ref{appendixB} for a definition. This factor is {\em essentially unique}, that is unique  modulo right multiplication by an arbitrary $(m\times m)$ constant orthogonal matrix.

The rank deficiency of the spectrum $\Phi$ and consequently of the process  $y$  appears in models used in a variety of applications and is discussed in the literature   from dif-ferent points of view.\\
Singular autoregressive (AR) or autoregressive moving average (ARMA) models are discussed in \cite{BLM19}, \cite{Deistler19}, \cite{Ferrante-20cdc}. These models  make contact  with   dynamic factor analysis representations; see \cite{DeistlerEJC}, \cite{BG-15} where an essential role is played by a rank-deficient component driven by the common factors. They occur in  biological networks reconstruction as  discussed in \cite{nwbio09}, \cite{Yuan-11}. Low rank processes are also encountered  in graphical models which are common in social networks, see \cite{Ferrante-18letters}, \cite{Ferrante-20tac}, \cite{Ferrante-21tac}, \cite{Zorzi-16}. Specific engineering examples where identification of rank-deficient processes is involved are discussed  in \cite{TR06},\cite{LSTB19}.\\
The identification  of  singular processes has recently been addressed in
\cite{ChiusoP-12}, \cite{VanDenHof17}, \cite{WVD18}, \cite{BLM19}, \cite{WVD18-2}, \cite{BGHP-17},  and  \cite{PCL-21}. Some of these papers, like \cite{VanDenHof17}, \cite{WVD18-2}, propose  an  ingenious  adaptation of  the Prediction Error Method (PEM) identification and are of special interest. We shall briefly comment on their approach later in this paper. For  a recent  survey of the literature see \cite{Lippi-D-A-22}.\\
Let the process $y$ be partitioned as
\begin{equation}\label{ysplit}
    y(t):=\bmat y_1(t) \\ y_2(t) \emat,
\end{equation}
where $y_1(t)$, $y_2(t)$ are jointly stationary of dimension $m$ and $p$. By properly rearranging the  components of $y$, we can assume that $y_1$ is a process of full rank $m$.  The spectral density can then be partitioned as
\begin{equation}
\label{Phidecomp}
\Phi(z)=\begin{bmatrix}
\Phi_{11}(z)&\Phi_{12}(z)\\ \Phi_{21}(z)&\Phi_{22}(z)
\end{bmatrix},
\end{equation}
where $\Phi_{11}(z)$ is full rank.\\
It is well-known that the PEM identification procedure requires that there must be  a unique representation of the predictor in terms of past $y$. This is not the case unless the minimum phase spectral factor $W(z)$ is square full rank. In fact, let  $e(t)$ be the $m$-dimensional normalized innovation of $y$ and let us expand the innovation representation $y(t)= W(z) e(t)$ as
$$
y(t)\!= [W(\infty)+ z^{-1} \hat W(z)]\, e(t)= W(\infty)e(t) +  \hat W(z) e(t-1),
$$
where the last term is a causal function of the strict past innovations and is therefore (by the well-known causal equivalence of a process and its innovation) must be the predictor, although  expressed as {\em a function of $e$}. Now since $W(z)$ is  not invertible,  there is no unique expression of $e(t)$ as a function of past $y$ and therefore there is no unique  expression of the predictor as a function of past $y$. Although $W(z)$ is full column rank, its left inverse is not unique, and one could end up with many expressions for the predictor. This difficulty is exacerbated when one is working with (parametric) estimates of the transfer function. Therefore a direct  application  of the PEM  principle seems to be forbidden due to the reduced-rank  noise.
However  in \cite{VanDenHof17}, \cite{WVD18-2} the authors  essentially  show that the past of the first component $y_1$ acts as a {\bf sufficient statistic} for the predictor so that  there is a unique  expression of the joint predictor which is a  function only of the past of $y_1$. This remarkable re-presentation unfortunately requires a crucial minimum phase condition which is not always satisfied.\\
In this paper we follow  a different approach based on ideas first presented in   \cite{CLPcdc20}, \cite{CLP-21} and especially in \cite{PCL-21}. In the early paper \cite{GLSampling} it  was shown   that there must exist a, in general non-causal,  deterministic relation between the components of a singular vector process $y$. In \cite{CLPcdc20}, \cite{CLP-21} and in \cite{PCL-21}  the existence and structure of such deterministic relations is elucidated and specified  as a component of a special feedback model for the joint process. \\
We should advise the reader that in the setting of this paper, the deterministic relation between the variables  $y_1(t)$ and $y_2(t)$, is in a sense ``dual" of  that introduced in  \cite{CLPcdc20} and also studied in  \cite{CLP-21}. This relation is described by a rational transfer function which  can be identified quite easily by a least squares algorithm.\\
The structure of this paper is as  follows. In Section~\ref{secFB} we introduce the feedback model representation of   low-rank processes  and prove the  existence of a deterministic dynamical relation which reveals the special structure of these processes. In Section~\ref{secIden} we exploit the special feedback structure for identification of the deterministic relation and of the transfer functions of the two stochastic components driven by white noise.
In Section~\ref{Ident}  we study the identifiability  of the transfer functions  of a  feedback representation. The feedback structure  is  in general not identifiable and  a characterization of all  equivalent forward loop transfer functions is provided based on classical result of stabilization theory in robust control. Even under the constraint of stability of the forward loop, yet there are  infinitely many equivalent (stable) forward transfer functions  which realize the same transfer function of the  feedback model.
The existence of a canonical (unique) pair of  transfer functions  of the feedback loop  is discussed in Subsection~\ref{stableWiener}. This canonical structure is a causal Wiener filter plus an orthogonal error term.
The identification of this canonical feedback structure is discussed in Section \ref{Wiener}. The canonical model  has an   output-error representation  where the additive error is not necessarily white. Two possible approaches to  the identification of this model are briefly discussed.\\
 From Section~\ref{secIden} to Section~\ref{Wiener} we discuss the identification of low rank time series.
The identification of processes with an external measurable input is  considered in Section~\ref{secIdenExInput}, where we also make a brief comparisons with the work of \cite{VanDenHof17}, \cite{WVD18-2}.
Several simulation examples are reported in Section~\ref{secExamples}. Finally, in Section~\ref{secCon} we come to some conclusions.\\
{\em Notation:} All random processes in this paper are  discrete-time ($t\in \Zbb$), wide sense stationary with zero mean and finite variance. Most notations comply with those used in the book \cite{LPBook} and should be quite standard in the system identification literature. In particular, multiplication by $z$ is the one step ahead shift operator acting as: $z y(t) = y(t+1)$ and $y(t)=W(z) u(t)$ designates  the response of a linear system with transfer function $W(z)$ to an input function $u\equiv \{u(t); t\in \Zbb \}$.  A rational vector or matrix   function is called {\em stable} if all of its poles  belong to the interior of the unit disk.  The strictly proper stable rational vector functions written as $n$-dimensional column vectors form a distinguished subspace of the vector Hardy space $H^2_n$ which, with some abuse of notation, in this paper will be denoted by the same symbol. $\bar H^2_n$ will denote the direct sum of $H^2_n$ plus the constants. This space contains the causal rational functions which are finite for $z\to \infty$ (but are not necessarily strictly causal). The  notation $[\cdot]_+$ stands for the orthogonal projection operator onto  $\bar H^2_n$. It should be remembered that it maps rational functions into {\em  proper} stable rational vector functions.

\section{Feedback models of stationary processes }\label{secFB}
\begin{figure}[h]
      \centering
      \includegraphics[scale=0.45]{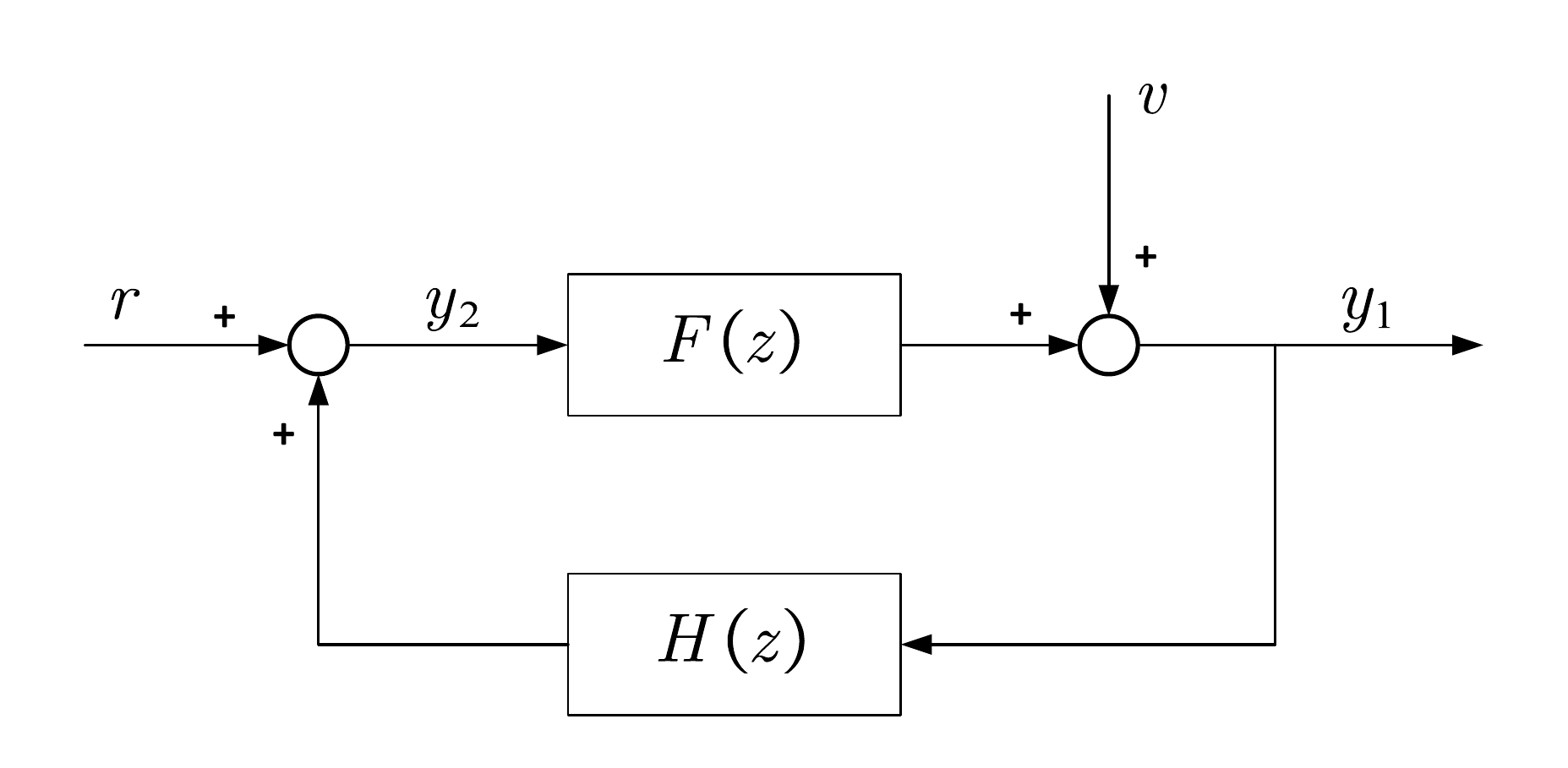}
      \caption{Block diagram illustrating a feedback model}
      \label{FigFH}
\end{figure}

In this section, inspired by   classical references such as  \cite{Caines-C-75} \cite{Gevers-A-81}, \cite{CainesBook} and \cite[Sect. 17.1]{LPBook}, we   review  the definition and some properties of general feedback models which have been also used in our recent papers \cite{CLPcdc20}, \cite{PCL-21} and \cite{CLP-21} in the context of rank-deficient vector processes. Then we  derive a special feedback model for low-rank processes and prove the existence of a deterministic relation between $y_1(t)$ and $y_2(t)$.
\begin{defn}[Feedback Model]\label{Def1}
A {\em Feedback model} of the  process $y(t):= \bmat y_1(t)^{\top} & y_2(t)^{\top}\emat^{\top}$ of dimension~$m+p$, is a pair of equations
    \begin{subequations}\label{fbmodel}
       \begin{align}
        \label{fby}
        y_1(t) &=F(z)y_2(t) + v(t),\\
        \label{fbu}
         y_2(t) &= H(z)y_1(t) + r(t), \quad t\in \mZ
    \end{align}
    \end{subequations}
satisfying the following conditions:
  \begin{itemize}
   \item $v$ and $r$ are jointly stationary uncorrelated processes called the {\em input noise} and the {\em modeling error}; 
     \item $F(z)$ and $H(z)$ are $m\times p$, $p\times m$ causal transfer function matrices, one of which is strictly causal, i.e., has at least one delay;
      \item the closed loop system mapping $\bmat  v\\ r\emat  \to  \bmat y_1\\ y_2\emat$  is well-posed and {\em internally stable };
        \end{itemize}
\end{defn}
 The block diagram illustrating a feedback representation is shown in Fig.~\ref{FigFH}. Note that the transfer functions  $F(z)$ and $H(z)$ are in general not stable, but the overall feedback configuration needs to be internally stable \cite[Chap. 3.2]{DFT}. In the sequel, we shall often suppress the argument $z$ whenever there is no risk of misunderstanding.
The following construction  shows that feedback representations of p.n.d. jointly stationary processes always exist. Let $\Hb_t^-(y_1)$ be the closed span of the past components $\{ y_{11}(\tau),  \dots, y_{1m}(\tau)\}\mid \tau \leq t\}$ of the vector process $y_1$ in an ambient  Hilbert space of second order zero-mean random variables \cite{LPBook} and let $\Hb_t^-(y_2)$ be defined likewise in terms of $\{ y_{21}(\tau),  y_{22}(\tau), \dots,  y_{2p}(\tau)\mid \tau \leq t\}$. A representation similar to \eqref{fbmodel}  may be gotten from the formulas for causal Wiener filters expressing both $y_1(t)$ and $y_2(t)$ as the sum of the best linear estimate based on the past of the other process plus an error term
\begin{subequations} \label{Wiener}
        \begin{align}
    y_1(t) &= \mathbb{E}\{y_1(t)\mid \Hb_{t-1}^-(y_2)\} + v(t), \\
    y_2(t) &= \mathbb{E}\{y_2(t)\mid \Hb^-_{t}(y_1)\} + r(t).
\end{align}
\end{subequations}
For a processes  with a rational spectral density the Wiener predictors can be expressed in terms of causal rational transfer functions $F(z)$ and $H(z)$ as in Fig \ref{FigFH}.
Here we choose $F(z)$ to be strictly causal. An alternative  representation with $H(z)$ strictly causal can also be given, to guarantee  well-posedness of the feedback system.
Although the errors $v$ and $r$ obtained by the procedure \eqref{Wiener} may be correlated, in Appendix~\ref{appendixA} we will show that there exist feedback model representations where they are uncorrelated. The following  theorem describes basic properties of feedback representations of stationary processes. It has been proven in \cite{CLPcdc20}, \cite{PCL-21} and is also reported in the companion paper \cite{CLP-21}, therefore its proof is omitted.
\begin{thm}
    \label{lem1}
        The transfer function matrix $T(z)$ from $\begin{bmatrix}v\\ r\end{bmatrix}$ to $\begin{bmatrix} y_1\\ y_2\end{bmatrix}$ of the feedback model is given by
\begin{subequations}\label{TPQ}
\begin{equation}  \label{T}
T(z)=\begin{bmatrix}P(z)&P(z)F(z)\\Q(z)H(z)&Q(z)\end{bmatrix},
\end{equation}
with
\begin{equation}  \label{PQ}
\begin{split}
P(z)&=(I - F(z)H(z))^{-1},    \\
Q(z)&= (I - H(z)F(z))^{-1}
\end{split}
\end{equation}
\end{subequations}
where the inverses exist. Moreover, $T(z)$  is a full rank (invertible a.e.) and (strictly) stable function which yields
\begin{equation} \label{Phi}
\Phi(z)=T(z)\begin{bmatrix}\Phi_v(z)&0\\0&\Phi_r(z)\end{bmatrix}T(z)^*,
\end{equation}
where $\Phi_v(z)$ and $\Phi_r(z)$ are the spectral densities of $v$ and $r$, respectively, and $^*$ denotes transpose conjugate.
\end{thm}
Since $T(e^{i\theta})$ has full rank a.e., $\Phi$ is rank deficient if and only if at least one of $\Phi_v$ or $\Phi_r$ is. Thus the rank of $\Phi$ is equal to the sum of the ranks of $\Phi_v$ and $\Phi_r$. The next lemma will play a crucial role in this paper. Although it can be seen as a dual of a continuous-time result in \cite{CLPcdc20}, for the benefit of the reader we shall provide a proof anyway.
\begin{lem}\label{lem2}
Suppose $(F\Phi_rF^*+\Phi_v)$ is positive definite a.e. on the unit circle. Then
\begin{equation} \label{H}
H=\Phi_{21}\Phi_{11}^{-1} - \Phi_rF^*(\Phi_v+F\Phi_rF^*)^{-1}(I-FH),
\end{equation}
that is
\begin{equation}\label{DefH}
H=\Phi_{21}\Phi_{11}^{-1}
\end{equation}
if and only if $\Phi_r \equiv 0$.
\end{lem}
\begin{pf}
From \eqref{TPQ} and \eqref{Phi}, we have
\begin{align*}
   \Phi_{21}&= Q(H\Phi_v +\Phi_r F^*)P^*=QH\Phi_vP^* + Q\Phi_rF^*P^*,  \\
   \Phi_{11}  &=  P(\Phi_v+F\Phi_rF^*)P^* ,
\end{align*}
and using the easily verified relations
\begin{displaymath}
PF=FQ, \quad HP=QH.
\end{displaymath}
we get
$
 \Phi_{21}=HP\Phi_vP^* +Q\Phi_rF^*P^*.
 $
 Adding and subtracting the term $HP F\Phi_rF^*P^*$ we end up with
\begin{align*}
 \Phi_{21}  &=  H \Phi_{11}+ (Q-QHF)\Phi_rF^*P^*\\
 & =H \Phi_{11} + \Phi_rF^*P^*
 \end{align*}
 since $Q-QHF=I$. Then  \eqref{DefH} follows if and only if  $\Phi_r=0$ since $P$ is invertible  and $F$ times a spectral density can be identically zero only if the spectral density is zero as otherwise  this would imply that the output process of a filter with stochastic input would have to be orthogonal to the input.
 \hfill$\Box$\end{pf}
In the following we specialize to feedback models of rank deficient processes. We shall show that there are  feedback model representations  where the feedback channel is described by  a {\em deterministic relation} between $y_1$ and $y_2$.
\begin{thm}\label{thmspecialfb}
Let $y$ be an $(m+p)$-dimensional process  of rank $m$. Any full rank $m$-dimensional subvector process $y_1$ of $y$ can be represented by a feedback scheme of the form
\begin{subequations}\label{fbr0}
 \begin{align}
        y_1 &= F(z)y_2+ v, \\
    y_2 &= H(z)y_1.
\end{align}
\end{subequations}
   where the transfer functions $F(z)$ and $H(z)$ satisfy the conditions of Definition \ref{Def1} and the input noise $v$ is of full rank $m$.
\end{thm}

\begin{pf}
Recall that $n$-tuples of real rational functions form a vector space $\mR^n(z)$ where the rank of a rational matrix is the rank almost everywhere. The claim is equivalent to the two statements\\
1. If we have the structure \eqref{fbr0}, i.e. $\Phi_r \equiv 0$;  then $y_1$ is   of full rank $m= \rank (\Phi)$.\\
2. Conversely if $y_1$ is   of full rank $m= \rank (\Phi)$ then $\Phi_r \equiv 0$.\\
Part 1 follows from Lemma \ref{lem2} since because of \eqref{Phi} then $\Phi_v$ must have rank $m (= \rank (\Phi))$.\\
Part 2 is not so immediate. One way  to show it could be as follows.\\
Since 	$\Phi(z)$ has rank $m$ a.e. there must be a full rank $p \times (m+p)$ rational matrix which we write in partitioned form, such that
\begin{align}\label{AB}
[A(z)~B(z)] \Phi(z) \!&=\!0\, \Leftrightarrow \, [A(z)~B(z)]\! \bmat\Phi_{11}(z)\\ \Phi_{21}(z)\emat \!=\!0
\end{align}
where $A$, $B$ are $p\times m$ and $p\times p$ matrices, which is equivalent to   $  A(z)y_1(t) +B(z)y_2(t) \equiv 0 $.\\
We claim that $B(z)$ must be of full rank $p$. One can prove this using the invertibility of $\Phi_{11}(z)$.
For, suppose $B(z)$ is singular, then  pick a $p$-dimensional non-zero row vector $a(z)$ in the left null space of $B(z)$ and multiply from the left the second relation by  $a(z)$. This would imply that also $a(z)A(z)\Phi_{11}(z)=0$ which in turn implies $a(z)A(z)=0$ since $\Phi_{11}$ is full rank. However $a(z)[A(z)~B(z)]$ cannot be zero, for the  matrix $[A(z)~B(z)]$ is full rank $p$ and hence $a(z)$ must be zero. So $B(z)$ must be full rank.\\
Now  take any nonsingular $p\times p$ rational matrix $M(z)$ and consider instead $M(z)[A(z)~B(z)]$, which  provides an equivalent relation to \eqref{AB}.
By choosing $M(z)=B(z)^{-1}$ we can reduce $B(z)$ to the identity to get
$$
[-H(z)~\, I\,] \bmat  y_1(t) \\ y_2(t)\emat =0
$$
where  $H(z)$ is a rational matrix function, so that    one gets the deterministic dynamical relation
$$
  y_2(t)= H(z) y_1(t)\,.
$$
Substituting in  the general feedback model one concludes  that $y_2(t)$ must then be a functional of only the noise $v$ since $y_1(t)$ is such. Therefore by the uncorrelation of $v$ and $r$ one must conclude that in the second equation of \eqref{fbmodel} $r$ must be  the zero process i.e. $\Phi_r=0$. Hence a representation like \eqref{fbr0} must hold.
\hfill$\Box$
\end{pf}
\section{Identification of low rank processes}\label{secIden}
Suppose we want to identify by a PEM method  a   model of an $(m+p)$-dimensional time series $y$ of rank $m$. To this purpose, the model class should be selected to guarantee   identifiablility (i.e. uniqueness) and it is specific of the PEM method that it should   actually be  an {\em innovation representation} of $y$ which is well known to be essentially unique. This representation involves a  {\em minimum phase} spectral factor $W(z)$ satisying  \eqref{sfd}  whereby
\begin{equation} \label{WNmodel}
 y(t) = W(z)e(t),
\end{equation}
where     $e(t)$ is the $m$-dimensional normalized innovation process of $y$, a white noise of covariance $I_m$. \\
Consider then the model \eqref{WNmodel} block-partitioned as in \eqref{ysplit},
\begin{equation}\label{W1W2}
   y(t)= \bmat y_1(t) \\ y_2(t) \emat := \bmat W_1(z)\\ W_2(z) \emat e(t),
 \end{equation}
where $y_1$ and $y_2$ are described by the special feedback model \eqref{fbr0}. From the defining property  of $y_1$ and $y_2$ in our partition,  $W_1(z)$ must be square $m\times m$, stable, causal and  non singular (invertible a.e.) and $W_2(z)$    stable and causal.
\begin{prop}\label{Hprop}
The transfer function of the feedback channel in model \eqref{fbr0} is given by the expression
\begin{equation}\label{HfromWs}
H(z) = W_2(z)W_1(z)^{-1}
\end{equation}
and is unique. In fact, it depends  only on the joint spectrum \eqref{Phidecomp}.  Stability of $H$ holds if and only if $W_1$ is minimum phase.
\end{prop}
\begin{pf}
The formula follows from the partition \eqref{W1W2} since both components are driven by the same full rank process $e(t)$.
 Formula \eqref{DefH} in Lemma~\ref{lem2},  provides the alternative expression $H(z)=\Phi_{21}(z) \Phi_{11}(z)^{-1}$ which must obviously coincide with  \eqref{HfromWs}  since $\Phi_{2,1}(z)=W_2(z)W_1(z)^{*}$ and $\Phi_1(z)=W_1(z)W_1(z)^{*}$. It is then clear that  $H(z)$ depends only on the joint spectrum \eqref{Phidecomp} and must therefore be {\em unique} for a given partition of the vector process $y$. That stability of $H$ holds if and only if $W_1$ is minimum phase follows since there cannot be cancellations in forming the quotient \eqref{HfromWs}. It is shown in Appendix \ref{appendixB} that if $W(z)$ is minimum phase then $W_2(z)$ and $W_1(z)$ cannot have common unstable zeros which could cancel in forming the product \eqref{HfromWs}. \hfill $\Box$
 \end{pf}

 \nd {\bf Remark:} Proposition~\ref{Hprop} is in agreement with  \cite{CLP-21}, where it was shown that $H(z)$, (called $F$ in [24]) is unique but in general not stable by a counterexample provided in Section V-A.  (Also see the conference version \cite{CLPcdc20}). Incidentally this answered a question by Manfred Deistler in the negative.
 On the contrary we shall see that there are in general infinitely many transfer functions $F(z)$  generating $y$ by means of the model \eqref{fbr0}.

\subsection{Estimation of $H(z)$}
Since the relation between $y_2$ and $y_1$ is completely deterministic  we can  identify  $H(z)$ by imposing  a deterministic transfer function model  to the observed data. The model can be  written  as $ A(z^{-1}) y_2(t) - B(z^{-1})y_1(t)=0,\, t=1,\ldots,N$ (the minus sign is for convenience) where  $A(z^{-1})$ and $B(z^{-1})$ are matrix polynomials in the delay variable $z^{-1}$, of dimension $p\times p$ and $p\times m$  such that
  $$
  H(z)= A(z^{-1})^{-1}B(z^{-1})\,.
  $$
 One can always choose $A(z^{-1})$ monic and parametrize the matrix polynomial  $B(z^{-1})$ so  that the transfer function corresponds to the   difference equation
\begin{equation} \label{procedure1}
y_2(t)= -\sum_{k=1}^q A_k y_2(t-k) + \sum_{k=0}^r B_ky_1(t-k), \quad t=1,\ldots,N,
\end{equation}
 where we have written  $A(z^{-1}) = I +\sum_{k=1}^{q}A_k z^{-k}$ and   $B(z^{-1})= \sum_{k=0}^{r}B_k z^{-k}$. The above equation  involves  delayed components of the  observed trajectory data of $y$. The  coefficients can then be estimated  by  solving a deterministic overdetermined  linear system by least squares and a strongly consistent and unbiased result can be obtained whether the system  is stable or not, assuming we know the true degrees of $A$ and $B$. See the example in subsection \ref{Ex3}. \\
 Then, once $W_1$ is identified, the transfer function $W_2$ can be calculated using the relation
 \begin{equation}\label{W2fromW1}
 W_2(z)=H(z)W_1(z).
 \end{equation}
 This procedure however may fail if the true $W_1(z)$ in \eqref{WNmodel} is not minimum phase  and the identification   is done by a time-recursive least squares algorithm.  In fact if $W_1(z)$ has  unstable zeros then $H(z)$ is unstable and  in this case the noise superimposed  to the data may tend to excite the unstable modes of the system  \eqref{procedure1} and cause divergence. To bypass  the constraint of minimum phase of the true system   one should rely on  algorithms processing the whole data batch in one shot.

 \subsection{Identification of $W_1$}\label{RecW}
Next, since $y_1$ (and $W_1$) is full rank, it seems that one could easily  identify, say an ARMA innovation model  for $y_1$ based only on observations of $y_1(t)$ on some large enough time interval. By this procedure   we would ideally identify an innovation  representation  for $y_1$, say   $y_1(t)=G_1(z)e_1(t)$ where however  the minimum phase transfer function $G_1(z)$ does not necessarily coincide with the upper block of the joint innovation representation of $y$. This would be true only if the upper
block $W_1(z)$ of the minimum phase $W(z)$ was also minimum phase, {\em which in general may not be true} (the same clearly holding also for the lower block). See Appendix \ref{appendixB} for a discussion of this point.
In other words,  the partitioned  innovation representation of the full process $y$ may not necessarily coincide with  the separate innovation representations  of the two components $y_1$ and $y_2$.\\
Therefore a PEM method applied to measurements of $y_1$ may not lead to a consistent estimate of the upper block $W_1(z)$ of the model \eqref{W1W2} since  there may be  a nontrivial inner function $Q_1(z)$ such that
\begin{equation}\label{OutInnW1}
W_1(z)= G_1(z) Q_1(z)
\end{equation}
One may then wonder if the identification problem we are after is well-posed and if there actually is a procedure to recover a non-minimum phase $W_1(z)$ from the data. To this end we shall first show  that estimating $G_1$  can   nevertheless lead to a consistent estimate of the joint spectrum.
 \begin{prop}
Assume that the transfer function $H(z)$ is  estimated as described in the previous subsection, that is using the data $(y_1,y_2)$ and asymptotically satisfying    the relation \eqref{W2fromW1}. Then, even if the  upper block $W_1(z)$ of the joint (minimum phase) transfer function $W(z)$ is not  minimum phase, a consistent  estimate of the minimum phase transfer function $G_1(z)$  does  nevertheless produce  a consistent estimate of the joint spectral density of the (joint) process $y$.
\end{prop}
\begin{pf}
 The statement is obviously true for the auto spectral density $\Phi_{11}(z)$. Then just recall  that the cross spectral density of $y_2$ and $y_1$ can be  expressed as
$$
\Phi_{21}(z)=  H(z) \Phi_{11}(z)=H(z)G_1(z) G_1(z)^{*}.
 $$
 Using the estimate $\hat G_1(z)$ in place of $W_1(z)$ in formula \eqref{W2fromW1} to compute the  estimate $\hat W_2(z) $,   although $\hat W_2(z):= \hat H(z)  \hat G_1(z) $ may be a non-consistent estimate of  $W_2(z)$, it does  result in a consistent estimate of the cross spectrum    $\Phi_{21}(z)$. A similar argument can be used for $\Phi_{22}$.  \hfill $\Box$
 \end{pf}
 Hence a consistent  estimate of the minimum phase transfer function $G_1(z)$  does   produce  a consistent estimate of the joint (minimum phase) transfer function $W(z)$   of the (joint) process $y$ and therefore also  of its   $m\times m$ upper  block.

\subsection{ Procedure  to recover $W_1$ and $W_2$ from consistent estimates of $G_1$ and $H$.
(Equivalently, recovering the missing  inner factor $Q_1$ in the outer-inner factorization \eqref{OutInnW1}).}
From the expression $ H(z)=W_2(z)Q_1^*(z)G_1(z)^{-1}$, that is from
\begin{equation}\label{CopInn}
    H(z)G_1(z)=W_2(z)Q_1^*(z)=W_2(z)Q_1(z)^{-1}\,.
\end{equation}
One can get  estimates of $W_2$ and $Q_1$  by performing a right-coprime factorization
in the rational $H^{\infty}$ space (see e.g. \cite[sect. 5.4]{Zhou-D-G-95}), of the  estimated product $\hat H(z)\hat G_1(z)$ imposing that $Q_1$ should be inner (see e.g. \cite{Oara-V-99}).  This guarantees uniqueness, see again \cite[p. 368]{Zhou-D-G-95}. The conjugate inner function $Q_1^*$ must contain exactly  all the unstable poles of the left member. \\
In this way we are in principle  able to obtain  a consistent estimate of the full minimum phase model $W$ even when $W_1$ is not minimum phase. The calculations are  easy when $W_1$ is  scalar but may be quite involved in the matrix case where one should  need to use coprime factorization algorithms in terms of state-space realizations which we shall not dwell into.

\section{Identification of the feedback model}\label{Ident}
The procedure described so far does not take into account the possibility of modeling the system by the structure \eqref{fbr0}, in particular by  the ``{\em internal}" feedback des-cription  of $y_1$ involving the transfer functions $F,\,K$ and $H$. Assume that the model \eqref{W1W2} is  in innovation form, with   $e(t)$ the  innovation of the joint process $y(t)$ and let
\begin{subequations}\label{y2Hy1}
\begin{align}
   y_1 &= F(z)y_2+K(z)e,    \label{y1Fy2Ke}         \\
   y_2 &= H(z)y_1.
\end{align}
\end{subequations}
be the corresponding feedback representation with $K(z)$  a square  spectral factor such that $v(t):=K(z)e(t)$, which we assume  minimum phase for  identifiability. From \eqref{TPQ} we have
\begin{equation}
\label{WPQK}
\bmat W_1 \\ W_2 \emat =T\bmat K\\ 0 \emat = \bmat PK \\ QHK\emat =\bmat PK \\ HPK\emat ,
\end{equation}
with both  $P(z)K(z)$ and $H(z)P(z)K(z)$    submatrices of a minimum phase transfer function.\\
One may ask how one could  recover the direct transfer function $F(z)$ from the identified $W_1(z)$ and $H(z)$. This would amount to solving for $F$ the relation $W_1=(I-FH)^{-1}K$ which, assuming $H$ is given, contains {\em two unknowns}. Hence $F(z)$ and $K(z)$ are  {\em not identifiable} as they do not correspond uniquely to the minimum phase representation \eqref{W1W2} and hence do not correspond uniquely to the joint spectral density of $y(t)$. In other words, there are in general infinitely many pairs $(F(z), K(z))$ realizing in feedback form the innovation representation \eqref{W1W2}. This actually agrees with the well-known identifiability analysis of feedback systems which dates back to \cite{Gustavsson-L-S-77}, see the example in Sect. VI.

\subsection{On equivalent feedback structures}\label{Stabiliz}
In our setting the causal  transfer  function $H(z)$ of  the feedback channel is uniquely determined by the two components of the process $y$, once the partition is fixed and known, while there are in general a multitude of pairs $(F,K)$ yielding the same transfer function $W_1(z)$. Note that each such pair should make $W_1$ stable. In particular, once $H$ is given, each $F$ should make  the feedback configuration \eqref{fbr0} internally stable. In this subsection we shall characterize the set of such equivalent $F$'s. This problem   can be regarded as the ``dual" of a stabilization problem in control, which is also discussed in our companion paper \cite{CLP-21} on modeling of low rank vector processes. Here we have a more limited scope than in \cite{CLP-21} as we only want to  analyze the identifiability of the system by explicitly describing all pairs  of transfer functions $(F,K)$ which realize  the same stable $W_1$.\\
Since the feedback system must be internally stable the sensitivity function $P(z)$ defined in \eqref{PQ} needs to be analytic in the complement of the open unit disk, without unstable pole-zero cancellation between $F(z)$ and $H(z)$.  Assuming for the moment that $H(z)$ is a proper stable rational function,   there is a whole class of proper rational  functions $F(z)$  which accomplish this job. In the scalar case they are all described  by the formula \cite[Chapter 5.1]{DFT},
\begin{equation} \label{Fstab}
F(z)= \Frac{S(z)}{1+S(z) H(z)}
\end{equation}
where $S(z)$ is an arbitrary proper stable rational function.   The corresponding sensitivity function is  given by
$$
P(z)= 1+S(z)H(z)
$$
linearly parameterized by an arbitrary such $S(z)$. All corresponding $K(z)$ are then obtained from the relation \eqref{y1Fy2Ke}, that is
$ K(z)= P(z)^{-1}W_1(z)$
so that  all such $(F,K)$ yield the same transfer function $W_1(z)$.\\
When $W_1(z)$ is not minimum phase and  $H(z)= W_2(z) W_1(z)^{-1}$ fails to be  stable,  closed-loop stability can still  be characterized by using a coprime stable proper-rational factorization of $H(z)$ yielding a more general parametrization of all $F$'s as described in \cite[Sect. 5.4]{DFT} (involving the so-called Youla parametrization).\\
In the matrix case, still assuming  a stable $H$, there is  a parametrization formula similar to \eqref{Fstab}, see e.g. reference \cite{Zhou-D-G-95}. But for the unstable case  one needs to  use matrix coprime factorizations to obtain the stabilizing  $F$. This issue is fully discussed in the dual context of our companion paper  \cite{CLP-21}.

\subsection{A canonical feedback model} \label{stableWiener}
As seen from \eqref{Fstab}, there are infinitely many possible transfer functions $F(z)$ (and also companion $K(z)$) realizing the same closed loop transfer function $W_1$.
In this subsection  we shall ask the following natural question: If one restricts   $F$  to be stable and causal, does there exists a {\em unique }feedback representation \eqref{y2Hy1}? Since the identifiability analysis of the previous section involves also the transfer function $K(z)$, it is quite evident that the answer should be negative. The following example provides in fact a few different pairs $(F,K)$, all with a strictly causal stable $F$,  which realize the same transfer function $W(z)$.

\begin{example}
Let a $2\times 1$ transfer function $W(z)$ be partitioned by two scalar blocks of respective transfer functions
\begin{subequations}\label{ex1W1W2}
\begin{align}
W_1(z) &=  \frac{z^3}{(z-0.5)(z+0.5)(z-0.2)}, \\
  W_2(z) &= \frac{z^3}{(z-0.5)(z-0.2)(z+0.1)}.
\end{align}
\end{subequations}
the corresponding transfer function $H$ being (from \eqref {HfromWs})
\begin{equation}\nonumber
 H(z)=\frac{z+0.5}{z+0.1}.
\end{equation}
We can provide three  different pairs  $F,K$ realizing the system, all three with a stable strictly causal $F$. The first being
\begin{equation}\nonumber
    F_1=\frac{-0.4}{z+0.5}, \quad
    K_1=\! \frac{z^3}{(z-0.5)(z-0.2)(z+0.1)}.
\end{equation}
the second,
\begin{equation}\nonumber
    F_2=\frac{0.4}{z+0.5}, \quad
    K_2=\! \frac{z^3(z-0.3)}{(z+0.5)(z-0.5)(z-0.2)(z+0.1)}.
\end{equation}
and finally
\begin{equation}\label{ex1F3K3}
    F_3=\frac{(0.2z^2+0.25z-0.5)(z+0.1)}{(z+0.5)z^3},\quad
    K_3=\! 1.
\end{equation}
To check that all three pairs realize the  minimum phase $W_1$ in the example, just calculate the noise transfer functions  $K_i$   from $K_i= (I-F_iH)W_1$, yielding all $K_i$ to be minimum phase, and the corresponding $P_i=(I-F_iH)^{-1}=W_1K_i^{-1}$ being stable.
\end{example}
This last example offers a hint  leading to the characterization of uniqueness: one can choose a particular function $F(z)$ which, besides being stable with at least  one unit delay, acts as the transfer function of the Wiener predictor of $y_1(t)$ based on the (strict) past of $y_2$. Then one should have a representation like \eqref{y1Fy2Ke} where $v(t)$ is the {\em prediction error}, uncorrelated with (i.e. orthogonal to) the past space $\Hb_{t-1}^-(y_2)$.\\
The proof of uniqueness of such a representation is just based on the  uniqueness of the orthogonal  decomposition of $y_1(t)$ as a linear causal functional of the strict past of $y_2$ plus an  error part orthogonal to the past space $\Hb_{t-1}^-(y_2)$. By the orthogonal projection lemma \cite[p. 27]{LPBook}, given such a decomposition, the linear causal functional of the strict past of $y_2$ must then be the (unique) orthogonal projection $\mathbb{E} [y_1(t)\mid \Hb_{t-1}^-(y_2)]$ onto $\Hb_{t-1}^-(y_2)$, i.e. the Wiener predictor.\\
In particular, when $K(z)$ is  a constant matrix as in the third example,  the noise $K_3 e(t)$ is automatically orthogonal to the strict past space of $y_2$ and we automatically get the  remarkable interpretation of $F(z)$ as the transfer function of the Wiener predictor. Indeed, below we shall show that  this will surely happen  when $W_2$ is minimum phase.

\begin{thm}\label{lemcalF}
 Assume that $W_2$ is minimum phase; then  there is a representation \eqref{y2Hy1}  where  $F$  is stable and strictly causal, that is
$F(z)=z^{-1}\bar{F}(z)$ with $\bar{F}(z)$ causal and stable (analytic in $\{|z| \geq 1\}$) and $K(z)$ is a constant matrix $K_+$. In fact, this
$\bar{F}(z)$
coincides with  the transfer function $F_+(z)$  of the  one-step ahead Wiener predictor based on the strict past of $y_2$,  that is
\begin{equation}\label{Fplus}
F_+(z) y_2(t-1) =\mathbb{E}\{y_1(t)\mid \Hb_{t-1}^-(y_2)\}
\end {equation}
and  the prediction error $\tilde y_1(t):= y_1(t)-F_+(z) y_2(t-1)$ can be written  $K_+ e(t)$ where $e(t)$ is the innovation of   the  joint process $y$.  The  representation
\begin{equation}\label{Feedplus}
y_1(t)= F_+(z) y_2(t-1) +K_+ e(t)
\end {equation}
 is  the unique feedback  representation of $y_1(t)$ in which $v(t)$ is uncorrelated with the strict past of $y_2$.
 \end{thm}
\begin{pf}
Let  $W_2(z)=G_2(z)Q_2(z)$ with $Q_2(z)$   the inner factor of $W_2(z)$; it is a standard fact explained for example in  \cite[Chap. 3]{LPBook} that the Fourier representative of $\Hb_{t-1}^-(y_2)$ is the subspace $Q_2H^2_m$ of $H^2_m$. Denoting by $P^{ Q_2H^2_m}$ the orthogonal projection operator onto $Q_2H^2_m$, we can   write  the formal representative of the  error process $\tilde{y}_1(t):= y_1(t)- \E [y_1(t)\mid \Hb_{t-1}^-(y_2)]$ as
$$
\tilde{y}_1:= W_1 e - [P^{ Q_2H^2_m}W_1] \,e
$$
so that
\begin{equation}\label{OrthK}
K(z):= W_1(z) - [P^{ Q_2H^2_m}W_1](z)
\end{equation}
is the transfer function of the error process $v(t):=K(z)e(t)$ which by construction  is uncorrelated with the strict past $\Hb_{t-1}^-(y_2)$. In other words,
\begin{equation}\label{Kperp}
K(z)\perp Q_2H^2_m
\end{equation}
the orthogonality being understood as holding columnwise in the $L^2$ space of vector functions on  the unit circle.\\
Now if (and only if) $Q_2(z)=I_m$ then $K(z)\perp H^2_m$ which means that $K(z)$ (in fact its column functions) belong to the orthogonal complement $(H^2_m)^{\perp}$. But since $K(z)$ is analytic, this can happen only when $K(z)$ is a constant matrix. \hfill$\Box$
\end{pf}
Naturally, for a general representation \eqref{y1Fy2Ke}    with a strictly causal $F$, the error process $v(t)$, given by  $v(t)=[W_1(z)  -F(z) W_2(z) ] e(t):=K(z) e(t)$ may not necessarily be orthogonal to the past of $y_2$.

\section{Structure and estimation of the predictor}\label{Wiener}
 Denoting for convenience the one-step ahead predictor $\mathbb{E}\{y_1(t)\mid \Hb_{t-1}^-(y_2)\}$ by the symbol $\hat y_1(t)$, we may  calculate $F_+$ by the Wiener predictor formula.  see e.g.  \cite[p. 105]{LPBook}. Introducing the cross spectral density of the processes $y_1(t)$ and $e_2(t-1)\equiv z^{-1}e_2(t)$, one has
\begin{equation}\label{Fplus}
   F_+  =[ \Phi_{\hat y_1,z^{-1}e_2} ]_+ G_2^{-L}=   [z W_1Q_2^{*}]_+ G_2^{-L}\,.
\end{equation}
where $[\cdot]_+$ denotes the (causal) orthogonal projection of a  function onto the complete $\bar H^2_m$  space, $G_2(z)$ is the minimum phase factor of $W_2(z)$,  $G_2^{-L}$ its  (Moore-Penrose) left inverse and $Q_2(z)$ the inner factor of $W_2$ so that the innovation of $y_2$ is $e_2(t)= Q_2(z) e(t)$.   Hence, if $W_2$ is minimum phase  the  above simplifies to
\begin{equation}\label{WfiltMinPhase}
   F_+(z)  = [zW_1(z)]_+ W_2^{-L}(z)\,.
\end{equation}
and   one gets $ \hat y_1(t+1)= F_+(z)y_2 (t)= [zW_1(z)]_+\,e(t)$ and
so, when $W_1$ is also minimum phase, $e(t)= W_1(z)^{-1}y_1(t)$ and
\begin{equation}\label{FplusSc}
F_+(z) y_2(t)= z [W_1(z)-W_1(\infty)] W_1(z)^{-1}y_1(t)
\end{equation}
 is exactly    the one-step Wiener predictor of $y_1(t+1)$ given its own past. This agrees with the  sufficient statistic role of the past of $y_1$  in the predictor formulas (12) and (13) of \cite{VanDenHof17}.

\subsection{Estimation of $F_+(z)$}\label{subSecEstF}
A conceptually simple way to estimate $F_+(z)$ is to resort to estimates of the transfer functions $H(z)$ and the minimum phase factor $G_1(z)$. From consistent estimates of  these functions one can  perform  the coprime factorization \eqref{CopInn} to obtain estimates of  $W_2(z)$ and of the inner factor $Q_1(z)$. To estimate $G_2(z)$ and compute the inner factor $Q_2(z)$ one can then  perform an outer-inner factorization on the estimate of $W_2(z)$, i.e.,
$$
    W_2(z)=G_2(z)Q_2(z).
$$
With these data one may in principle compute $F_+(z)$  by formula \eqref{Fplus} and the companion noise transfer function $K(z)$ by implementing the formula \eqref{OrthK} or by $K=(I-z^{-1}F_+H)W_1$.
Although this may look like a rather complicated indirect procedure,  for scalar transfer functions it can be implemented quite easily, see Example~2 in Section~\ref{Ex3}.

 One may instead attempt to estimate  the transfer function $F_+(z)$ directly from the data.  For simplicity we shall restrict to the case of scalar processes, the generalization to the vector  case being relatively straightforward. We assume a rational structure, say
 $$
  F(z)= D(z^{-1})^{-1}N(z^{-1})
  $$
where $D(z^{-1})$ and $N(z^{-1})$ are polynomials in the delay variable $z^{-1}$, of degree $n$ and  $m$.
  Choosing $D(z^{-1})$ monic and  the numerator polynomial  $N(z^{-1})$ with a zero constant term,  the transfer function corresponds to the   difference equation
\begin{equation} \label{procedure1}
\hat{y}_1(t)= -\sum_{k=1}^n D_k \hat y_1(t-k) + \sum_{k=1}^r N_ky_2(t-k) \quad t=1,\ldots,N,
\end{equation}
involving  delayed components of the  unobserved trajectory of the predictor  $\hat y_1$ and of the `` input" time series $y_2$. Assuming we know the true orders, this could act as a parametric representation of the  predictor transfer function. Of course  $\hat{y}_1$ is not observed and the identification  problem needs to be formulated in an {\em output-error} setting.  Introducing  the prediction error
$$
v (t):= y_1(t)-\hat y_1(t)
$$
and  letting\\
$ \!\!\!\!\varphi(t-1) : = \!\bmat y_1(t-1) &...&y_1(t-n)&y_2(t-1) &... & y_2(t-r) \emat ^{\top}$\\
$
= \bmat \yb_1(t-1)\\ \yb_2(t-1) \emat\,,$
 where the boldface symbols $\!\yb_1(t-\!1)$, $\yb_2(t-1)$ represent arrays made  of $n$- and $r$-dimensional delayed variables $y_1(t-k)$  and $y_2(t-k)$ as specified by the model \eqref{procedure1}, the  representation \eqref{y1Fy2Ke} can be written as a ``constrained" pseudo-linear structure
\begin{equation}\label{OEmodel}
 y_1(t)= \varphi(t-1)^{\top} \theta +\varepsilon(t)\,.
\end{equation}
where $\theta$ is the $(n+r)$-dimensional vector of unknown parameters and $\varepsilon(t):= D(z^{-1}) v(t)$ still dependent on the parameter $\theta$. From what we have seen previously, in general $v(t)$ and hence $\varepsilon(t)$ may be  far from being white so attempts to use ARX identification may lead to badly biased estimates. In addition,  for $F_+(z) y_2(t-1)$ to be the Wiener predictor, $v(t)$ must be orthogonal to the strict past of $y_2$ which should  be added as a further constraint to the model.\\
The output-error model \eqref{OEmodel} could   be identified by an instrumental-variable method see \cite[p. 192-198]{Ljung}.  In the standard procedure the unknown  parameters   should first be roughly estimated   by     minimizing the average squared prediction  error $v(t)$ i.e. minimizing
 $$
 J_N(\theta):= \frac{1}{N}\sum _{t=1}^N v(t)^{2}
 $$
 by least squares pretending $v$ is white, that is imposing  orthogonality to the delayed data $\varphi(t-1)$, i.e.
 \begin{equation}\label{Orth}
  \frac{1}{N}\sum _{t=1}^N \varphi(t-1)  \,v (t) = 0\,.
\end{equation}
 which leads to the normal equations
 $$
  \frac{1}{N}\sum _{t=1}^N \varphi(t-1) \varphi(t-1)^{\top} \, \theta= \frac{1}{N}\sum _{t=1}^N \varphi(t-1)  y_1(t)\,.
 $$
 In the limit for $N\to \infty$ we are led to
 solve an equation of the form
\begin{equation}\label{Homogsol}
\Hb\, \theta=\E [\varphi(t) y_1(t)] \,.
\end{equation}
where  the matrix
$$
\Hb =  \bmat \Sigma_{\yb_1} &\Sigma_{\yb_1,\yb_2} \\
				\Sigma_{\yb_2,\yb_1} &  \Sigma_{\yb_2}	\emat
$$
 is formed by  obvious limit  covariance matrices of the observed data.
Due to the non-identifiability caused by  the deterministic feedback $y_2(t)= H(z) y_1(t)$ the matrix $\Hb$  turns out to have a large nullspace and the minimization does  not  lead to  a unique   estimate. Most standard software can however  compute a solution via the Moore Penrose pseudoinverse. A constraint which should be satisfied in order to get a consistent estimate of the transfer function $F(z)=z^{-1}F_+(z)$ is   the  stability of the estimated $D(z)$ polynomial.  This condition can   be imposed by implementing  a  spectral factorization procedure by which    the  estimated    parameters $D_k$ are substituted  by a spectrally  equivalent stable set via a fast Cholesky spectral factorization algorithm due to \cite{Bauer-55},\cite{Rissanen-73}. \\
The estimated $\hat D_k$ (and $\hat D(z)$) can then be used to filter the prediction error to improve the output error estimate   obtained as  $\hat\varepsilon(t):= \hat D(z^{-1}) v(t)$ and thereby implement an  iteratively  refined parameter estimation algorithm  by solving a sequence of {\em weighted} least squares problems.
 We shall however leave the analysis of this procedure   to a future publication.\\
As a simpler alternative, assuming $W_2$ minimum phase,   one may revert to   the simpler model \eqref{Feedplus} which  is    unique and hence identifiable and therefore
a Prediction Error method should be able to identify the  transfer function directly from observed data,  \cite[p. 203 ]{Ljung}.  One   may attempt a simple least squares estimation method by using a rational (or matrix-fraction) descriptions and transforming \eqref{Feedplus}  to  a constrained output-error model with a white output error. It is well-known that this model leads however  to a predictor which is a {\em nonlinear function of the parameters of the denominator} and the estimation procedure needs to be carried on iteratively. Moreover  the estimate is still  {\em constrained by the stabi-lity condition on} $F_+(z)$.  The naive  least squares method   can be consistent only if $F_+(z)$ is a FIR-type transfer function, that is the denominator of $F_+(z)$ is a constant (see again \cite[Sect. 7.3]{Ljung}). As a first approximation one  may use models of this kind.
  With this proviso, in spite of feedback, a suitably constrained  PEM method may work anyway \cite[p. 416]{Soderstrom-S-89}, \cite{Ljung}.

\section{Identification of a low rank model with an external input}\label{secIdenExInput}
\vspace{-0.5 cm}
Suppose we want to identify a multidimensional system  with an external input $u(t)$, say
\begin{equation}\label{WithInput}
   y(t)= F(z)u(t) + K(z) e(t)
\end{equation}
where $e$ is a white noise process. The input $u$ is assumed to be completely uncorrelated with $e$ (no feedback) and persistently exciting of an appropriate order. When $\dim e =\dim y$ and $K(z)$ is square invertible, one could attack the problem by a standard PEM method. The method however runs into difficulties when the noise  is of smaller dimension than $y$ since, exactly for the same reasons explained in Sect. \ref{Introd}, the predictor and the prediction  error are not well-defined.\\
When the dimension of $e$  is  strictly smaller than the dimension of $y$  the  model \eqref{WithInput}  is also called {\em low-rank}. This low-rank problem is  actually the one     discussed in \cite{VanDenHof17}, \cite{WVD18} and  \cite{WVD18-2} where the authors propose an approximate solution depending on a regularization parameter. In this section we shall propose a two-stage scheme to compute estimates of $F$ and $K$ which in principle does not use approximations.\\
Referring to the general feedback model for the joint process we can always assume $F$ causal and    $K(\infty)$ full rank and  normalized in some way. Consider then the prediction error of $y(t)$ given the past history of $u$. We have
\begin{equation}\label{tildey}
\tilde y(t):= y(t)- \E [y(t) \mid \Hb^-_t(u) ] = K(z) e(t)
\end{equation}
since, by causality of $F(z)$, the  Wiener predictor is exactly $F(z)u(t)$. Hence $\tilde y$ is a low rank time series in the sense described in the previous sections (now with the current $K(z)$ playing the same role of $W(z)$). In principle we could then  use the procedure described above for time series as we could preliminarily estimate $F(z)$  by solving a deterministic regression of $y(t)$ on the past of $u$ and hence get $\tilde y(t)$. If we choose linear least square methods, we will obtain a consistent estimation. Then a standard  ARMA identification can be applied to estimate the minimum phase  $K(z)$ in
terms of the pre-processed data  $\tilde y(t)$.\\
Compared with the approach in \cite{VanDenHof17}, \cite{WVD18} and \cite{WVD18-2}, we use a composition of basic least squares and ARMA identification methods which avoids the approxima-tions, and the possible complex computations of a regularized optimization problem with a tuning parameter.

\section{Simulation Examples}\label{secExamples}

\subsection{Example 1 [Both $W_1$ and $W_2$ minimum phase] }
\begin{figure}
      \centering
      \includegraphics[scale=0.6]{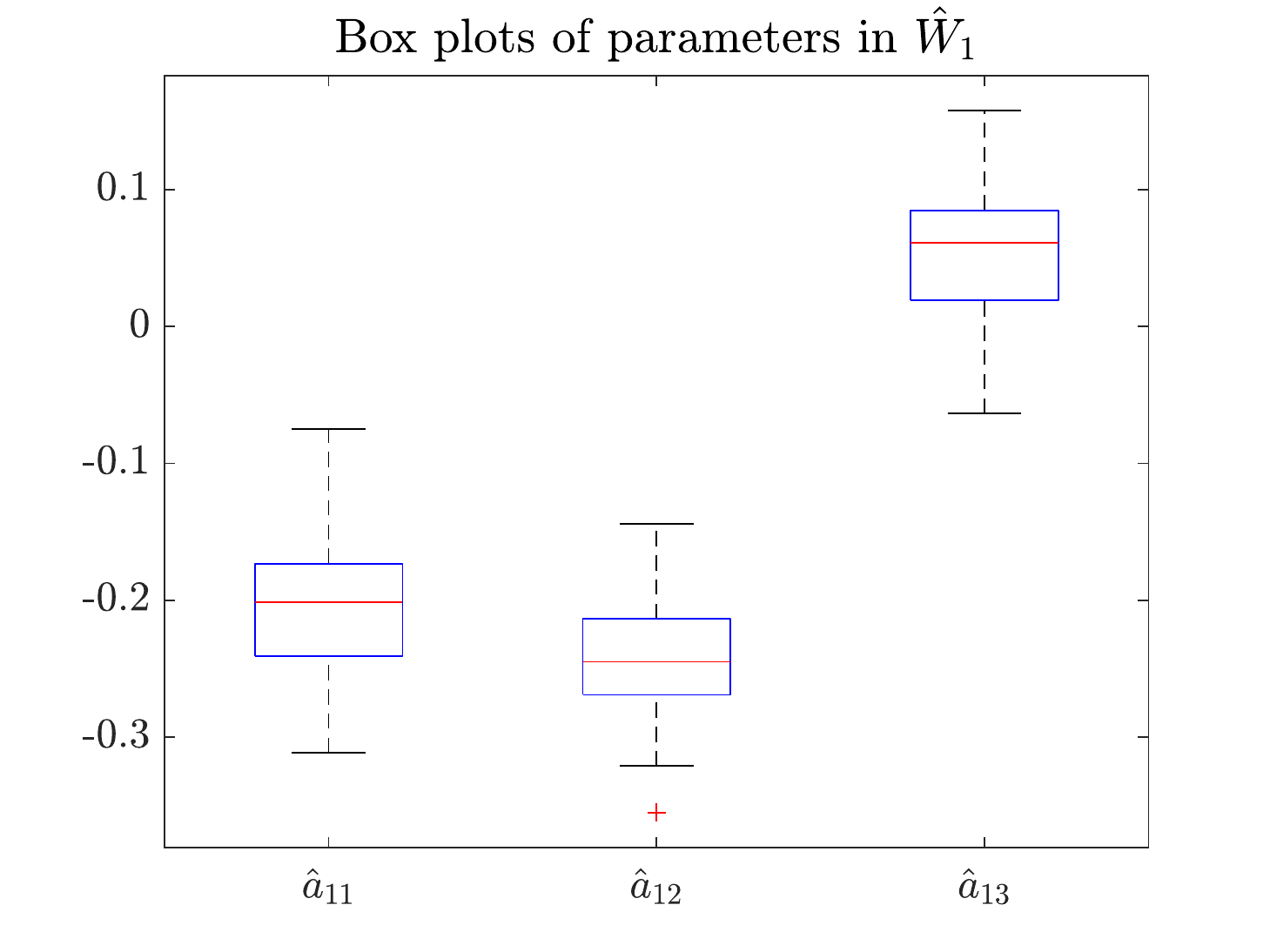}
      \caption {Box plots of $\hat{a}_{1k}$ for $k=1,2,3$ in example 1, where the true values are $a_{11}=-0.2,~a_{12}=-0.25,~a_{13}=0.05$.}
      \label{FigExmp1W1hat}
\end{figure}

\begin{figure}
      \centering
      \includegraphics[scale=0.6]{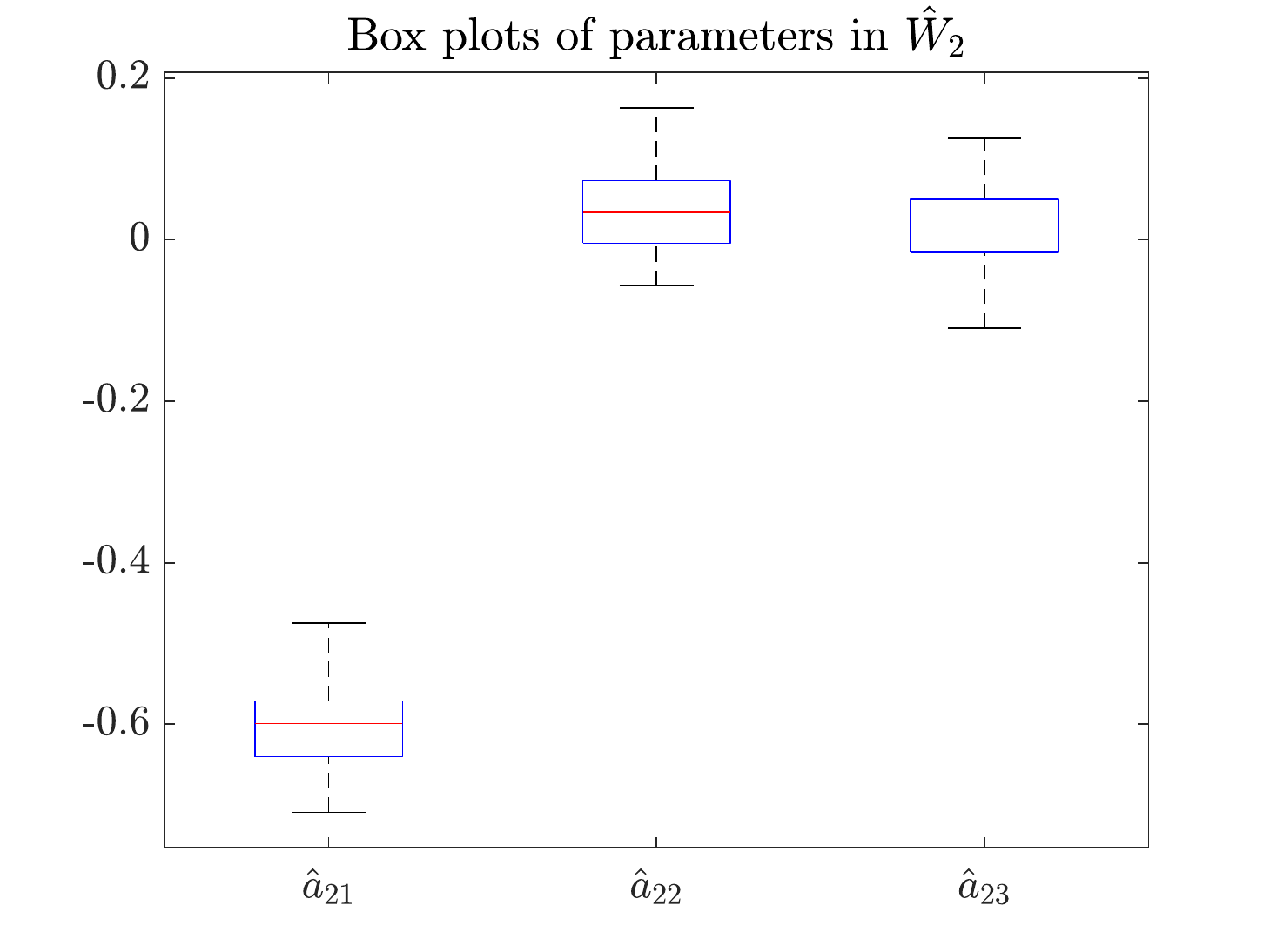}
      \caption {Box plots of $\hat{a}_{2k}$ for $k=1,2,3$ in example 1, where the true values are $a_{21}=-0.6,~a_{22}=0.03,~a_{23}=0.01$.}
      \label{FigExmp1W2hat}
\end{figure}

As a first simulation example  consider   a two-dimensional process of rank 1 described by
\begin{equation}\label{BothMinPhase}
 y(t)= \bmat W_1(z) \\ W_2(z) \emat e(t)
\end{equation}
where both $W_1(z)$ and $ W_2(z)$ {\bf are minimum phase rational transfer functions} and $e$ is a scalar Gaussian white noise of zero mean and variance $\lambda^2$. By simulation we produce a sample of two-dimensional output data of the system \eqref{BothMinPhase}. With these data we shall:
 \begin{itemize}
 \item Identify $W_1$ and $W_2$ by two separate AR models.
 \item Identify a transfer function model for $y_1$  and estimate $H(z)$ according to the first procedure described in Section \ref{RecW}. And then do the same for the other component.
  \item Estimate $F_+(z)$ and $K_+(z)$ in \eqref{Feedplus} using the estimated value of $W_1(z)$ and $H(z)$.
   \end{itemize}
We choose $W_1$ and $W_2$ as in \eqref{ex1W1W2} and  $e$   a scalar zero mean white noise of variance $\lambda^2= 1$. The process $y(t)$ has rank 1.  The two transfer functions  functions  $W_1$ and $W_2$  are normalized at infinity and minimum phase rational transfer functions. Note that in this particular example   both $y_1$ and  $y_2$ are full rank so that our procedure would work for both.\\
We have generated $100$ samples of the two-dimensional time series with $N=500$ data points $\{y_i(t); t=1,\ldots,N,\, i=1,2 \}$ and used   Monte-Carlo simulations in MATLAB. The results are condensed in  Box plots.\\
Assume the orders of $W_1$ and $W_2$ are known. Since the two AR models of $y_1$ and $y_2$ are  of order 3, we just implement two AR identification in MATLAB  for models of the form
$$
y_i(t) = -\sum_{k=1}^3 a_{i,k} y_i(t-k) + e(t), \qquad t=1, \ldots N,
$$
The box plots of the estimated parameters in $\hat{W}_1$ and $\hat{W}_2$ are shown in Fig.~\ref{FigExmp1W1hat} and Fig.~\ref{FigExmp1W2hat}. \footnote{In all box plots, the red horizontal line is the median of the data, the blue box contains half of the data points, the horizontal lines are at 25\% and 75\% level. The black tails (black horizontal lines) are at the minimum and maximum values, except for the outliers that are indicated by a red `$+$' sign.} In the two box plots, all median estimated values are close to the real ones, with the ranges of estimation values acceptable and only one outlier for $\hat{a}_{12}$.
We also use the average of 100 runs of Monte-Carlo simulation to estimate the asymptotic covariance of the estimated parameters which are of the order of  magnitudes $10^{-4}$, quite small compared with the magnitude of parameters. The  box plots in Figure \ref{FigExmp1W1hat} and \ref{FigExmp1W2hat} show that our AR estimators work well.

Next we do least-squares estimation of the transfer function $H(z)$. Since $H$ satisfies the identities
$$
    W_2(z)=H(z)W_1(z),\quad W_1(z)={\bar H}(z)W_2(z),
$$
 we can use  the following theoretical formula for $H$ and $\bar H$:
\begin{equation}\nonumber
 H(z)=\frac{1+0.5z^{-1}}{1+0.1z^{-1}},\quad {\bar H}(z)=\frac{1+0.1z^{-1}}{1+0.5z^{-1}}.
\end{equation}
which is equivalent  to the difference equation
\begin{equation}\nonumber
    (1+0.1z^{-1})y_2(t)= (1+0.5z^{-1})y_1(t),
\end{equation}
This is just a theoretical model which we keep for comparison.\\
Assuming now that we don't know the true degrees of the model polynomials in \eqref{procedure1}; then we   first carry on an order estimation to choose the appropriate $q$ and $r$ in the model
\begin{equation}\nonumber
  y_2(t)-y_1(t)= -\sum_{k=1}^{q}a_ky_2(t-k)+\sum_{k=1}^{r}b_jy_1(t-j),
\end{equation}
and then use least square  to get estimates of the parameters of the model
\begin{equation}\nonumber
   \hat H(z)=\frac{1+\sum_{k=1}^r \hat b_{k}z^{-k}}{1+\sum_{k=1}^q \hat a_{k}z^{-k}}.
\end{equation}
From a BIC table values we see that  when $(q,r)=(1,1)$ the BIC index reaches a minimum. So we do  least squares estimation of a first order  model
$$
 y_2(t)-y_1(t)=-a_1y_2(t-1)+b_1y_1(t-1).
$$
All the parameter estimates   turn out to be  equal to the true values of the parameters $a_1=0.1$, $b_1=0.5$, affected by extremely small errors.  In Monte-Carlo simulations, the calculated estimated variances are all smaller than $10^{-29}$. We don't show box plots here. For estimating $\bar{H}(z)$, we obtain very similar results, which are therefore not  presented. Here both $H$ and $\bar{H}$ are stable functions. We shall check if our algorithm also works when $H$ is not stable in the next example.

\begin{figure}
      \centering
      \includegraphics[scale=0.6]{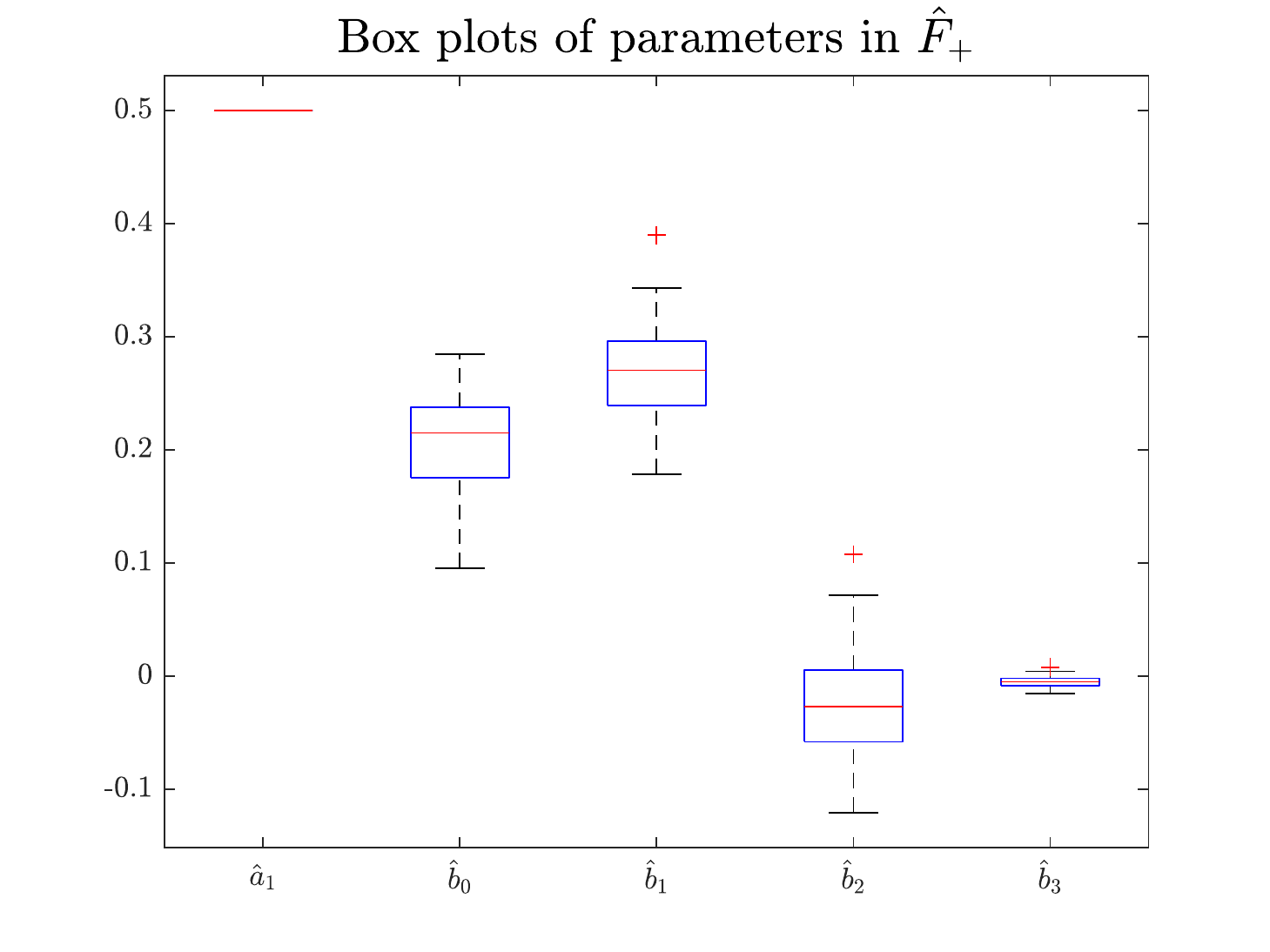}
      \caption{Box plots of the parameters in $\hat{F}_+$ in example 1.}
      \label{FigExmp1F+hat}
\end{figure}

Next we shall use the previous estimates $\hat{W}_1$ and $\hat{H}$ to calculate  estimates  of $F_+$ and $K_+$. We choose one estimate from  the previous  Monte-Carlo simulations, namely
\begin{equation}\nonumber
\begin{split}
   \hat{W}_1= &  \frac{1}{1 -0.1627 z^{-1} -0.2256 z^{-2} +  0.0505z^{-3}},\\
    \hat{H}=& \frac{1+0.5000z^{-1}}{1+0.1000z^{-1}}.
\end{split}
\end{equation}
From Theorem~\ref{lemcalF}, we know that there is one and only one pair  ($F_+, K_+)$ with $F_+$ the one-step Wiener predictor filter. In our case  $W_1$, $W_2$ are both normalized and minimum phase, and from \eqref{FplusSc}  we obtain the estimate of $F_+$ described by
\begin{equation}\nonumber
\begin{split}
    \hat{F}_+  &=z(1-\hat{W_1}^{-1})\hat{H}^{-1}\\
    &= \frac{(0.1627+0.2256 z^{-1}-0.0505z^{-2})(1+0.1000z^{-1})}{1+0.5000z^{-1}}
\end{split}
\end{equation}
and $\hat{K}_+$ equal to the constant part of $\hat{W}_1$, i.e.,
$$
    \hat{K}_+=\hat{W}_1(\infty)=1.
$$
The parameters of these functions are very  close to the true values
and hence appear to be  consistent estimates of $zF_3$, $K_3$ in \eqref{ex1F3K3}. \\
In fact, we get $\hat{K}_+=1$ each time in different simulations. What's more, since we are identifying with true orders in the previous Monte-Carlo simulations, we  obtain  a $\hat{F}$ with true orders as in \eqref{ex1F3K3}, i.e.,
\begin{equation}\nonumber
    F_+ =zF_3=\frac{0.2 + 0.27 z^{-1} - 0.025 z^{-2} - 0.005z^{-3}}{1+0.5z^{-1}}
\end{equation}
The box plot of the estimated parameters in $\hat{F}_+$, represented as
$$
    \hat{F}_+=\frac{\sum_{k=0}^3 \hat{b}_{k}z^{-k}}{1+\hat{a}_1z^{-1}}.
$$
 are  in Figure~\ref{FigExmp1F+hat}, showing that the estimate $\hat{F}_+$ obtained from $\hat{W_1}$ and the calculations in Section~\ref{Wiener} is a good estimate of the true causal Wiener filter $F_+$.

\subsection{Example 2 [Both $W_1$ and $W_2$  not minimum phase]}\label{Ex3}
In this subsection, a simple simulation example will be presented to show that our method {\bf can identify $H$ well also when it is unstable, can recover the minimum phase factor $G_1$ when $W_1$ is not minimum phase} as discussed in subsection~\ref{RecW}, and {\bf can estimate the Wiener Filter $F_+$ when $W_2$ is not minimum phase} as explained in the beginning of subsection~\ref{subSecEstF}.

Consider a two-dimensional process $y(t)$ described by \eqref{W1W2}, where  $e$ is a zero mean white scalar noise of variance $\lambda^2= 1$, and $W$ has the two blocks with transfer functions
\begin{equation}\nonumber
  W_1= \frac{z+2}{z-0.2}, ~~
  W_2=\frac{z-2}{z-0.2}.
\end{equation}
It is easy to obtain an outer-inner factorization of $W_1$ as in \eqref{OutInnW1}, where
\begin{equation}\nonumber
    G_1=\frac{2z+1}{z-0.2}=\frac{2+z^{-1}}{1-0.2z^{-1}},~~
  Q_1=\frac{z+2}{2z+1}.
\end{equation}
From these we  get the transfer function
$$
    H=\frac{1-2z^{-1}}{1+2z^{-1}},
$$
which is not stable. \\
Here for simplicity, we do not use Monte-Carlo simulations and order estimations. We just generate one group of data as in Example 1, with $e$ scalar zero mean and of variance $1$. Assume the orders of $G_1$ and $H$ are known.

\begin{figure}
      \centering
      \includegraphics[scale=0.6]{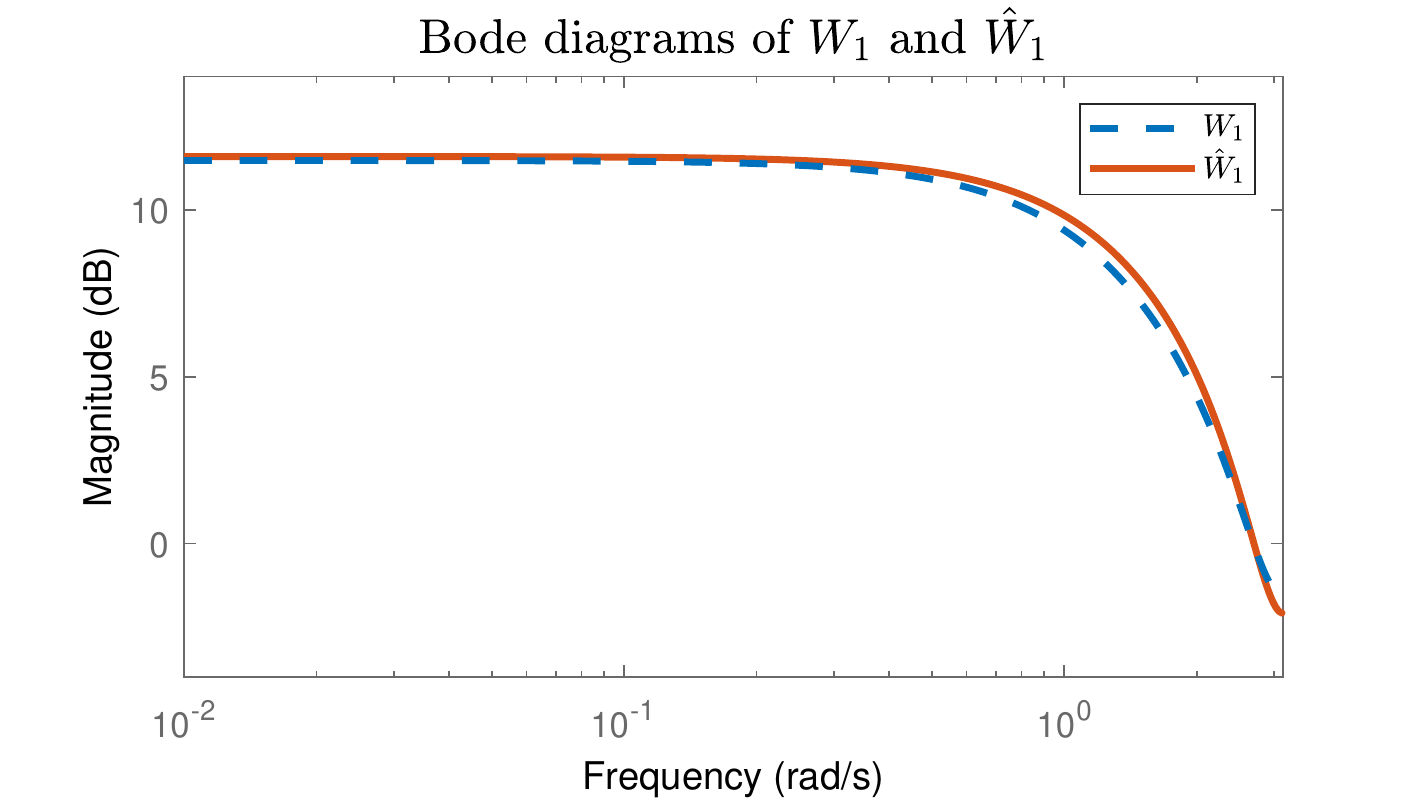}
      \caption {Bode diagrams of $W_1$ and $\hat{W}_1$ in example 2.}
      \label{FigExmpHuns}
\end{figure}

Though $G_1$ is not normalized at infinity, we may still implement an ARMA estimation  in MATLAB and obtain an estimated model
$$
 y_1(t)- 0.1442y_1(t-1)= \hat{e}(t)+ 0.5666\hat{e}(t-1),
$$
where the variance of the innovation $\hat{e}$ is $\hat{\lambda}^2=4.3127$. Then calculate the corresponding estimate of $G_1$
$$
    \hat{G}_1= \frac{\lambda(1+0.5666z^{-1})}{1-0.1442z^{-1}}=\frac{2.077 z + 1.177}{z - 0.1442},
$$
which is minimum phase.\\
Next we estimate $H$ by least squares on the model
$$
  y_1(t)+a_1y_1(t-1)=y_2(t)+b_1y_2(t-1),
$$
and obtain the estimate
$$
    \hat{H}=\frac{1+\hat{b}_1z^{-1}}{1+\hat{a}_1z^{-1}}=\frac{1-2.000z^{-1}}{1+2.000z^{-1}},
$$
which is practically equal to the true $H$, with an  estimation error variance of $1.0607\times10^{-29}$. Incidentally; all of our  simulation results show that the least squares method works well in identifying unstable $H$'s.\\
Since in this example $G_1$ and $W_2$ are scalar, we do not need coprime factorization for obtaining $Q_1$. In this case, $Q_1^*$ is the  conjugate inner factor of $H$ of formula \eqref{CopInn}, i.e., $Q_1$ is the greatest inner factor of $H^{-1}$. From
$$
 \hat{H}^{-1}=\frac{z+2.000}{z-2.000}=\frac{2.000z+1}{z-2.000}\cdot\frac{z+2.000}{2.000z+1},
$$
we have the estimate,
$$
 \hat{Q}_1=\frac{z+2.000}{2.000z+1}.
$$
Hence the estimate of $W_1$ is
$$
 \hat{W}_1=\hat{G}_1\hat{Q}_1=\frac{1.039 z^2 + 2.666 z + 1.177}{z^2 + 0.3558 z - 0.0721},
$$
whose magnitude Bode graph is compared with the true $W_1$ in Fig.~ \ref{FigExmpHuns}. The Bode diagrams show that we can obtain a consistent estimate of $W_1$ even if it is not minimum phase.
The corresponding  estimate of $W_2$ can be calculate from
$$
    \hat{W}_2=\hat{H}\hat{W}_1=\frac{1.039z^2- 1.489 z - 1.177}{z^2 + 0.3558 z - 0.0721},
$$
whose Bode diagram is close to that of the true $W_2$. We omit the graphs due to space limitations. It is easy to check $\hat{W}=[\hat{W_1}~\hat{W_2}]^{\top}$ is minimum phase.\\
Next we perform an outer-inner factorization on $\hat{W}_2$, i.e., $\hat{W}_2=\hat{G_2}\hat{Q}_2$, and obtain
$$
 \hat{G}_2=\frac{2.077z^2+ 0.1385 z - 0.5885}{z^2 + 0.3558 z - 0.0721},~
 \hat{Q_2}=\frac{z-2.000}{2.000z-1}.
$$
At last, the estimate of $F_+$ can be calculated by \eqref{Fplus}
$$
 \hat{F}_+=[z \hat{W}_1\hat{Q}_2^{*}]_+ \hat{G}_2^{-1}
 =\frac{0.3915z(z+0.6023)}{z^2+0.0667z-0.2834},
$$
and the companion noise transfer function $K(z)$ by implementing the formula \eqref{OrthK},
\begin{equation}\nonumber
\begin{split}
    \hat{K}& =\hat{W}_1 -z^{-1}[z \hat{W}_1\hat{Q}_2^{*}]_+\hat{Q}_2\\
 &=\frac{1.039z^3+1.7397z^2+0.4125z-0.0988}{z^3-0.1442z^2-0.2500z+0.0361}.
\end{split}
\end{equation}
It can be checked that $\hat{K}$ satisfies the equation $\hat{W}_1=(1-z^{-1}\hat{F}_+\hat{H})^{-1}\hat{K}$.
And we can see that when $W_2$ is not minimum phase, $K$ is not a constant anymore as stated in Theorem~\ref{lemcalF}.

\subsection{Example 3 [With external input]}
In this subsection we consider the identification of a two-dimensional process of rank 1 {\bf subjected to an external input } $u$. We generate a scalar white noise $u$ independent of $e$ and identify a 2-dimensional process model \eqref{WithInput} as described in the previous section \ref{secIdenExInput}.\\
In this example  the true system  is described by
\begin{equation}\label{Ex2Syst}
\begin{split}
   F(z) =&  z^{-1}\bmat 0.3 + 0.7z^{-1}+0.3z^{-2}\\
                 0.15 + 0.9z^{-1}-0.5z^{-2}\emat, \\
   K(z) =:& \begin{bmatrix}
              K_1(z) \\
              K_2(z)
            \end{bmatrix}= \bmat \frac{1+0.1z^{-1}+0.4z^{-2}}{1+0.3z^{-1}+0.4z^{-2}}\\ \frac{1+0.1z^{-1}+0.4z^{-2} }{1-0.2z^{-1}+0.1z^{-2}} \emat\,.
\end{split}
\end{equation}
where we have  used the same $F$ as in \cite{VanDenHof17} (called $G(q)$ there). Since  the $K_2$ of \cite{VanDenHof17} is not normalized to $1$,  we use a different one. Both components of our $K(z)$  here are normalized and minimum-phase  so the overall model is an innovation model. By calculation   the deterministic relation from $K_1(z)$ to $K_2(z)$ is
$$
    H(z)=K_2(z)K_1(z)^{-1}=\frac{1+0.3z^{-1}+0.4z^{-2}}{1-0.2z^{-1}+0.1z^{-2}}.
$$
For the model \eqref{Ex2Syst} we generate $100$ groups of two-dimensional time series of $N=500$ data points $\{y_i(t); t=1,\ldots,N,\, i=1,2 \}$. The Monte-Carlo simulations  are run with $u$ and $e$  independent scalar white noises of variances $2$ and $1$.
Of course here  we also measure  the input time series $u$. Suppose we do not know the  orders of both $F_i$'s  for $i=i,2$.

\begin{figure}
      \centering
      \includegraphics[scale=0.6]{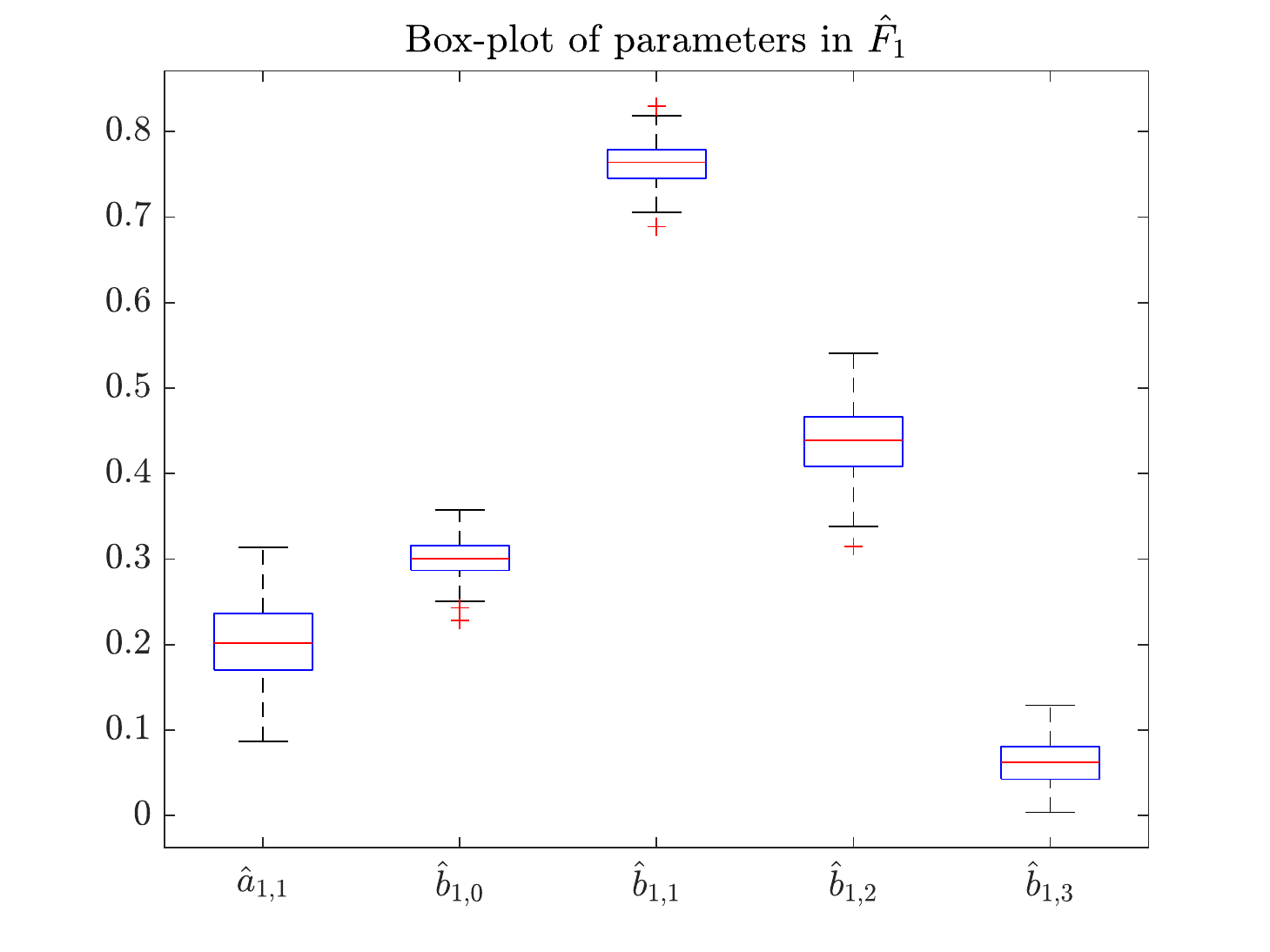}
      \caption {Box plots of parameters of $\hat{F}_1(z)$ in example 3.}
      \label{FigExmp2F1hat}
\end{figure}

\begin{figure}
      \centering
      \includegraphics[scale=0.6]{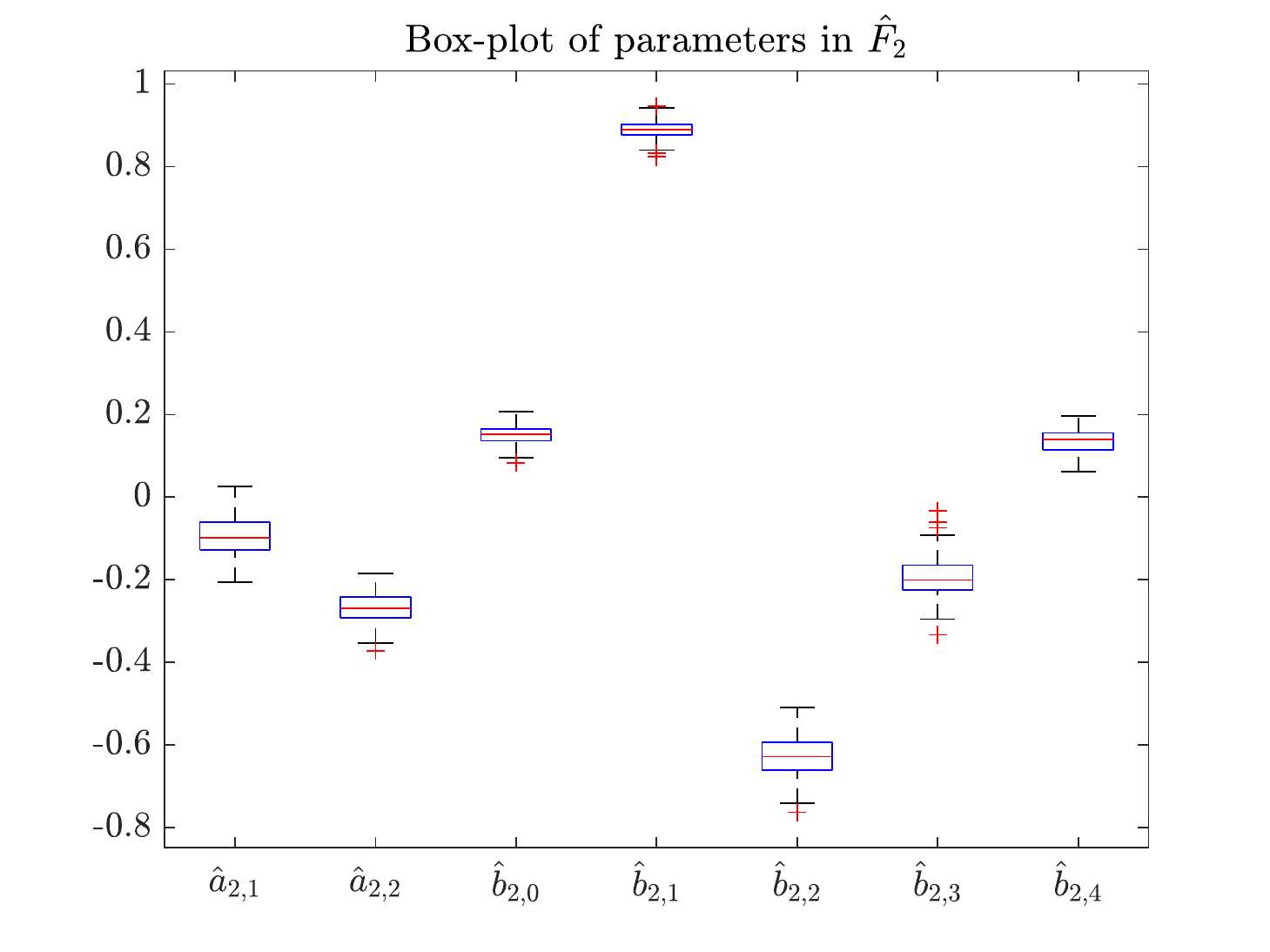}
      \caption {Box plots of parameters in $\hat{F}_2(z)$ in example 3.}
      \label{FigExmp2F2hat}
\end{figure}

First, let  $F_i(z)=z^{-1}A_i(z^{-1})^{-1}B_i(z^{-1})$ for $i=1.2$, where  the polynomials are parametrized as
\begin{equation}\nonumber
\begin{split}
  A_1(z^{-1})&=1+\sum_{k=1}^{q_1}{a}_{1,k}z^{-k},\quad
  A_2(z^{-1})=1+\sum_{k=1}^{q_2}{a}_{2,k}z^{-k}.\\
  B_1(z^{-1})&=\sum_{k=0}^{r_1}{b}_{1,k}z^{-k},\quad
  B_2(z^{-1})=\sum_{k=0}^{r_2}{b}_{2,k}z^{-k}
\end{split} \end{equation}
    corresponding  to the dynamic relations
\begin{equation}\nonumber
    A_i(z^{-1})y_i(t)=B_i(z^{-1})u(t-1)+\varepsilon_i(t),\quad  t=1,\ldots,N,
\end{equation}
$\; i=1,2$ where we have added a small white noise error term.
 We do a standard least squares  regression on these models, written in the form,
\begin{equation}
    \hat{y}_i(t)=-\sum_{k=1}^{q_i}{a}_{i,k}y_i(t-k)+\sum_{k=0}^{r_i}{b}_{i,k}u(t-1-k), \quad (i=1,2).
\end{equation}
where the orders are to be estimated. Order estimation by minimum  BIC leads to   choose $(q_1, r_1)=(1,3)$ and $(q_2, r_2)=(2,4)$. Although we don't get  the right model structures, with these orders we get the reasonable box plots shown in Fig.~\ref{FigExmp2F1hat} and Fig.~\ref{FigExmp2F2hat}, with very few outliers.

 Next we compute the deviations \eqref{tildey} by
$$
    \tilde{y}_i(t)= y_i(t)- \hat{F}_iu(t),\qquad i=1,2
$$
which are components of a 2-dimensional  low rank process. With these data  we estimate $K_1(z)$ and $K_2(z)$ by the  procedure  illustrated in Section~\ref{secIden}. This time, to smooth  the influence of   the wrong model structure used  in  estimating $F_1$ and $F_2$, we assume that  the true degrees of $H$ and $K_1$ are  known.

\begin{figure}
      \centering
      \includegraphics[scale=0.6]{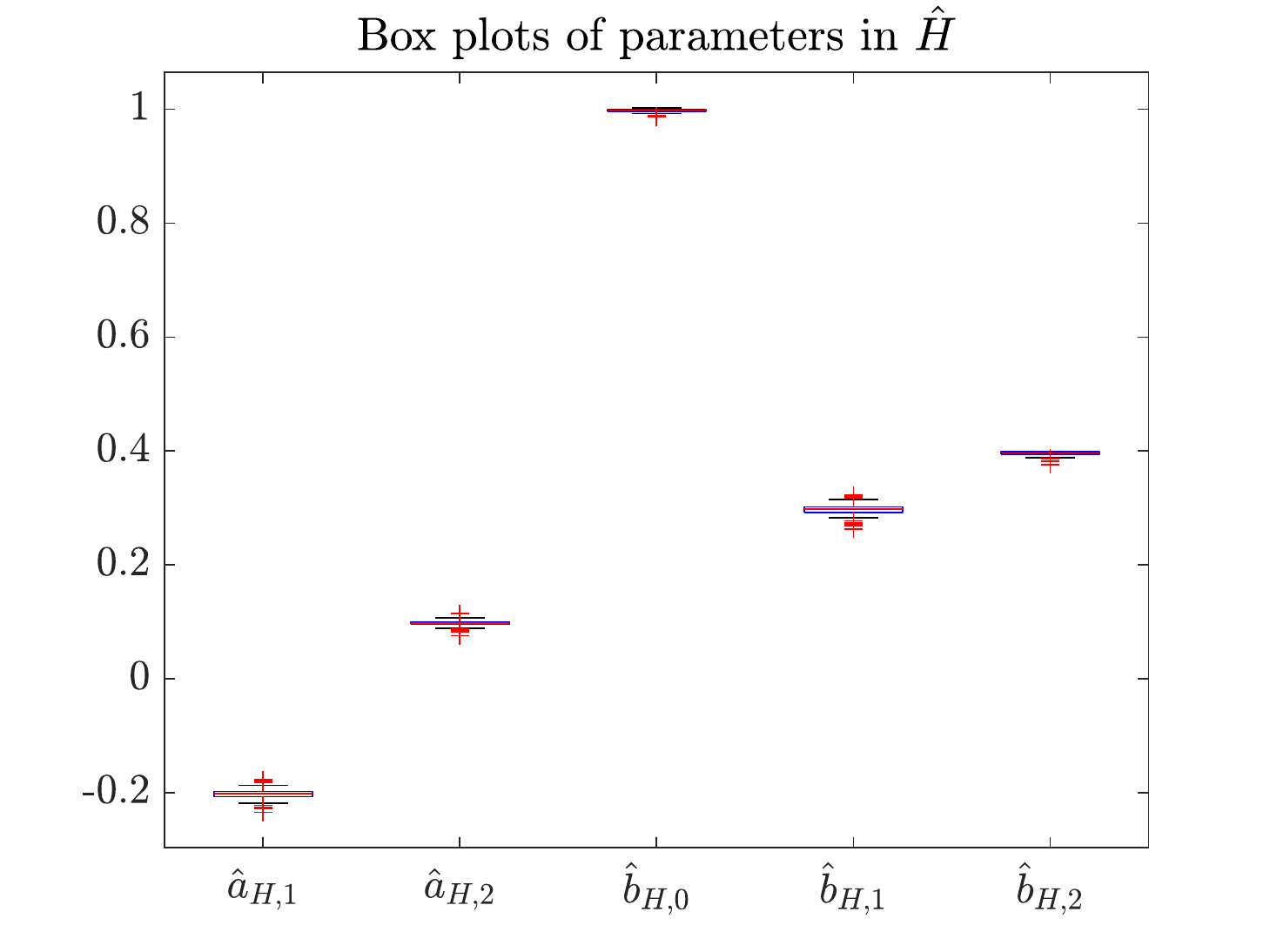}
      \caption {Box plots of parameters in $\hat{H}(z)$ in example 3.}
      \label{FigExmp2Hhat}
\end{figure}

\begin{figure}
      \centering
      \includegraphics[scale=0.6]{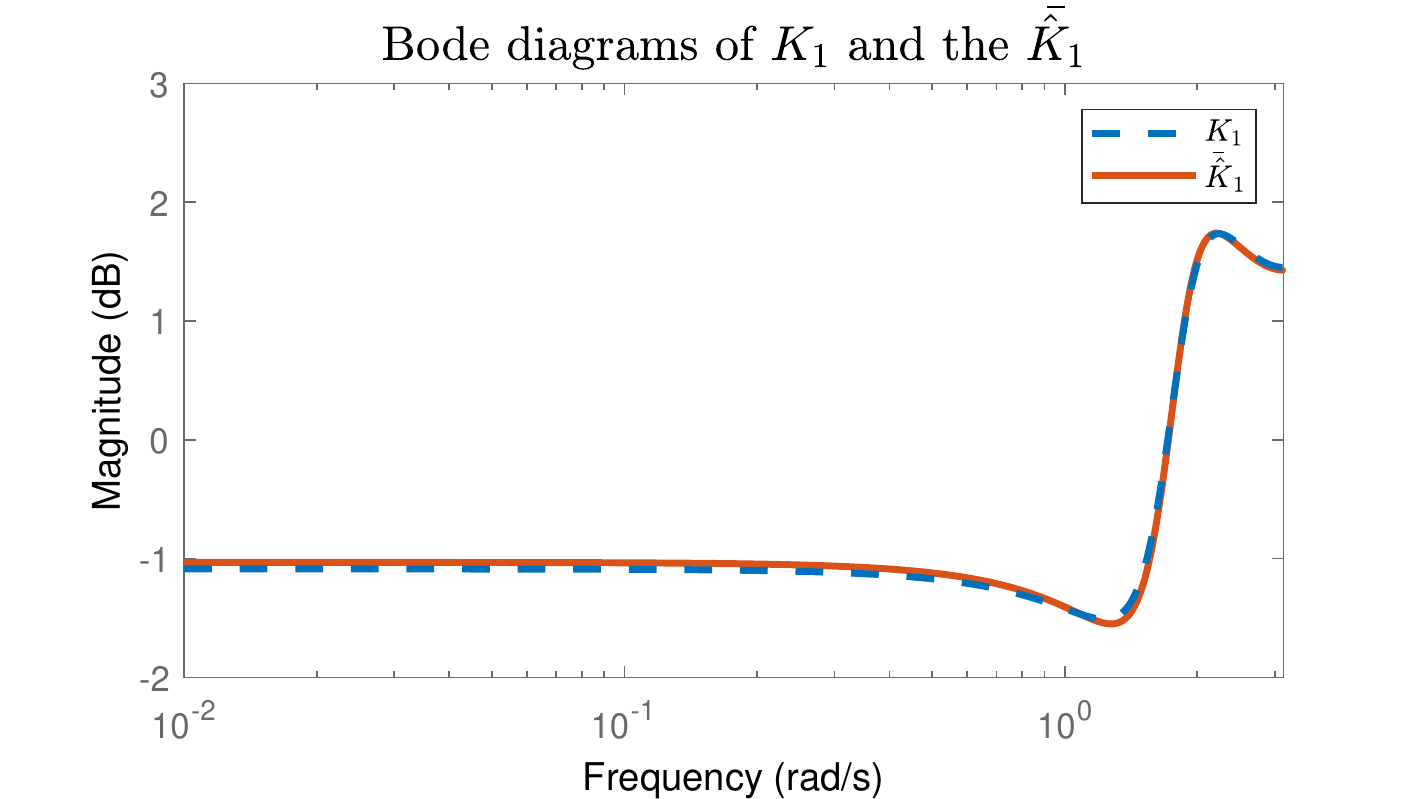}
      \caption {Magnitude Bode plots of   $\bar{\hat{K}}_1(z)$ of example 3.}
      \label{FigExmp2K1hat}
\end{figure}

\begin{figure}
      \centering
      \includegraphics[scale=0.6]{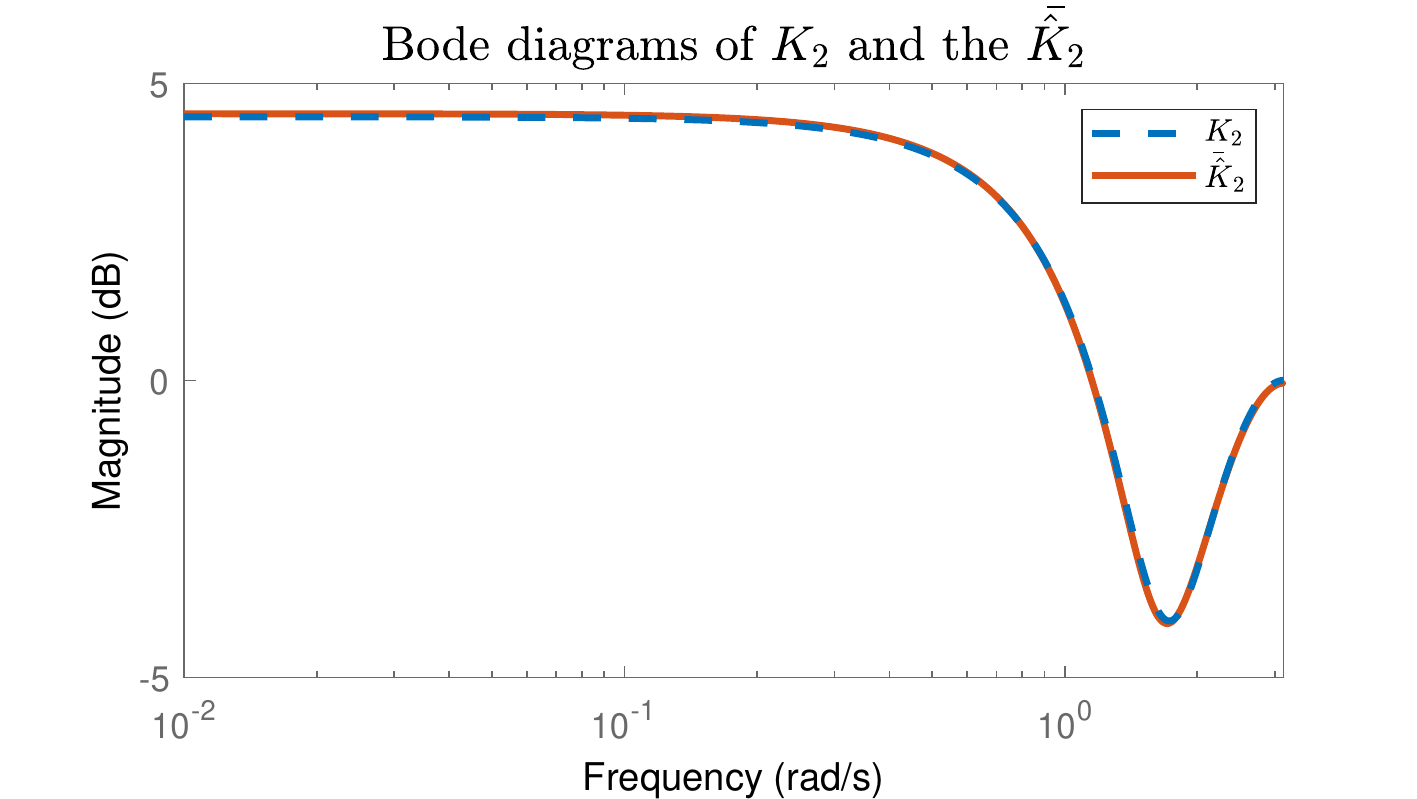}
      \caption {Magnitude Bode plots of  $\bar{\hat{K}}_2(z)$ of example 3.}
      \label{FigExmp2K2hat}
\end{figure}

 We first   use a least square method to estimate $H$ based on  the data   $\tilde{y_1}$ and $\tilde{y_2}$, assuming  true orders,
\begin{equation}\nonumber
   \hat H(z)=\frac{\hat b_{H,0}+\hat b_{H,1}z^{-1} + \hat b_{H,2}z^{-2}}{1+\hat a_{H,1}z^{-1}+\hat a_{H,2}z^{-2}}.
\end{equation}
Then let $K_1=A_1^{-1}C_1$ so that
\begin{equation}\nonumber
  A_1(z^{-1})\tilde{y}_1(t)=C_1(z^{-1})e(t),
\end{equation}
where
\begin{equation}\nonumber
\begin{split}
   A_1(z^{-1})&=1+a_{1,1}z^{-1}+a_{1,2}z^{-2},\\
   C_1(z^{-1})&=1+c_{1,1}z^{-1}+c_{1,2}z^{-2}.
\end{split}
\end{equation}
The box plot of the Monte-Carlo simulations of the estimates of $H(z)$ are shown in Fig.~\ref{FigExmp2Hhat}.
With the  estimate of $H(z)$  we can calculate the estimate of $K_2$ by
$$
    \hat{K}_2=\hat{H}\hat{K}_1.
$$
Because of multiplication of estimates, $\hat{K}_2$ turns out to have  a large number of parameters. In order to save space, we do not show their box plot.
Instead of drawing box plots, we have  compared the average of Monte-Carlo estimates with the true functions. Denote by $\bar{\hat{K}}_i$ $(i=1,2)$ the Monte-Carlo averages of the estimates $\hat{K}_i, i=1,2$;  the  Bode diagrams of the comparisons are shown in Fig.~\ref{FigExmp2K1hat} and Fig.~\ref{FigExmp2K2hat}. Both average estimates have Bode diagrams quite close to those of the true ones.
The results are nice even if we didn't guess the true model structures when estimating $F$.

\section{Conclusions}\label{secCon}
In this paper we have shown that a rank-deficient process admits a special feedback representation with a deterministic feedback channel, which  can be used to split the identification in two steps, one of which can be  based on standard PEM algorithms while the other is based on a deterministic least squares fit. Identifiability of these feedback structures is not  guaranteed and we show how to choose an identifiable representative. A consequent method of identifying low rank processes with an external input is also proposed.
It is shown that standard identification algorithms can be easily applied to identify the transfer functions of  low-rank models in diverse circumstances. Several simulations confirm the validity of the proposed approach.

\bibliographystyle{plain}        
\bibliography{auto21}           

\appendix
\section{Proof of the existence of models~\eqref{fbmodel} with uncorrelated noises}\label{appendixA}    
 Consider a feedback model like \eqref{Wiener} where the input noises $(r, v)$ may be correlated and let
 $$\hat r (t) :=\E[r(t) \mid v(s); s \in \mZ]
 $$ be the {\em acausal} Winer estimate of $r(t)$ given the whole history of the process $v$ \cite[p. 105]{LPBook}. Since the joint spectral density is rational there is a rational transfer function say $S(z)$ by which we can represent $\hat r$ as $\hat r (t)=S(z) v(t)$ (with the usual convention on  the  symbols). Hence
 $$
 r(t)= S(z)v(t) + w(t)
 $$
 where $w(t)$ is a stationary process uncorrelated with the whole history of $v$. Now, after substituting into the first equation,  the second equation of \eqref{fbmodel} can be written
 $$
 y_2(t)= [H(z)+S(z)] y_1(t) -S(z)F(z)y_2(t) +w(t)
 $$
 from which
\begin{align}
 y_2(t) &= [I +S(z)F(z)]^{-1} [H(z)+S(z)] y_1(t) \notag\\
 &+[I +S(z)F(z)]^{-1}w(t)
 \end{align}
 which, after setting $\tilde r(t):=[I +S(z)F(z)]^{-1}w(t)$ may be written $y_2(t)= \tilde H(z)  y_1(t) +\tilde r(t)$, of the same form of the second equation in \eqref{fbmodel} but now with $v$ and $\tilde r$  completely uncorrelated. \hfill $\Box$

  \section{On minimum phase matrix functions}\label{appendixB}
 Let $W(z)$  be an  $(m+p) \times m$ full column rank stable matrix possibly a spectral factor of our $(m+p) \times (m+p) $ spectral density matrix $\Phi(z)$ of rank $m$. Minimum phase functions are called {\em outer} in the mathematical literature. Although our functions are rational it will be convenient to refer to the general  definitions in Hardy spaces of the literature. For these we shall use the row-vector convention of the book \cite {LPBook}. The following is an intuitive definition which matches that for scalar functions \cite[Theorem 4.6.11, p.137]{LPBook}.

\begin{defn} \label{MinPhase} A rational matrix function
  $W(z)$ is minimum-phase, i.e., outer , if and only if it has all its poles in the open unit disc and all its zeros in the closed unit disc.
\end{defn}
One should refer to the definition of (right) zeros \cite[Definition 4.6.10, p.136]{LPBook} for full column rank  matrix functions with rows  in $H_{m}^2$.
For example, $\alpha$ is a zero of a $2\times 1$ matrix $W=[W_1, W_2]'$, if and only if it is a common zero of both $W_1$ and $W_2$. Equivalently there is a scalar inner function $q(z)$, a Blaschke product with a zero in $\alpha$, such that  $W(z)= \hat W(z) q(z)$ with $\hat W(\alpha)\neq 0$. More generally, we want to consider a  partition of $W(z)$
\begin{equation}\label{Partition}
    W(z)=\begin{bmatrix}
        W_1(z) \\
        W_2(z)
      \end{bmatrix}
\end{equation}
where $W_1(z)$, $W_2(z)$ are $m\times m$, $p\times m$ analytic matrix functions with rows in $H^2_{m}$. Next we recall the classical definition of an outer matrix function in the matrix Hardy space $H_{(p+m),m}^2$.
\begin{def}\label{outer}
The matrix function  $W(z)\in H_{(p+m),m}^2$ is  outer, if the row-span
$$
\overline{\Span}\,\{ \phi(z) W(z)\,;\, \phi \in H_{(p+m)}^{\infty}\}
$$
is the whole space $H_{m}^2$.
\end{def}
This is equivalent to saying that in the outer-inner factorization  $W(z)=\hat W(z)Q(z)$, the inner (matrix) function $Q$ must be a unitary constant, which we may identify with the the identity $I_m$.\\
Consider now the outer-inner factorizations
\begin{equation}
  W_1(z) = \hat{W_1}(z)Q_1(z),  \quad  W_2(z)  = \hat{W_2}(z)Q_2(z),
\end{equation}
where $\hat{W_1}$, $\hat{W_2}$ are the outer (minimum-phase) factors and  $Q_1, Q_2$ are inner (in fact matrix Blaschke products). The question we want to answer is: if $W$ is outer, does it follow that any (or both) of the two components $W_1, W_2$ should also be outer? We shall see that the answer is in general negative.\\
Let us recall the definition of {\em greatest common right  inner divisor} of two inner functions $Q_1$  and $ Q_2 $, see \cite[p. 188 top]{Fuhrmann-86} denoted $Q_1\wedge_R Q_2$. This is the inner function representative of the closed vector sum $H^2_{m}Q_1\vee H^2_{m}Q_2$.
\begin{thm}
Let a full column rank matrix function  $W(z)\in H_{(p+m),m}^2$ be partitioned as in \eqref{Partition}. The $W$ is outer if and only $Q_1$  and $ Q_2 $ are right-coprime, equivalently,  the {\em greatest common right  inner divisor} of $Q_1$  and $ Q_2 $  is the identity, i.e. $Q_1 \wedge_R Q_2=I_m$.
\end{thm}
\begin{pf}
 Follows from the identity see \cite[p. 188 top]{Fuhrmann-86}.
$$
H^2_{m}Q_1\vee H^2_{m}Q_2=  H^2_{m} (Q_1 \wedge Q_2)
$$
\end{pf}
  Hence $W(z)\in H_{(p+m),m}^2$ can be   outer even if  none of the two  submatrices $W_1$ and $W_2$ is. They just need to have no (unstable) zeros in common. On the other hand, when $W_1$ or $W_2$ have no unstable zeros, they are automatically outer.
\vspace{-2.mm}

\section{Stability of the Moore-Penrose Pseudo-Inverse} \label{MPinverse}
\vspace{-2.mm}
This  section was contributed by Augusto Ferrante \cite{Augusto-22}. It deals with stability of the Moore-Penrose pseudo inverse of a minimum phase rational matrix function. We might only concentrate on stability of a left inverse which is  what is needed in this paper but the result is more general.\\
Suppose we have a rational spectral factor $W(z)$ with $n$ rows and $p\leq n$ columns and   assume the minimum phase condition that $W(z)$ has full column rank for any $|z|\geq 1$.
This means that the {\em Smith-McMillan form} \cite[p.443-445]{Kailath-80}  of $W(z)$ has  the structure
$
W(z)=U(z) G(z) V(z)
$
where:\\
1) $U(z)$ is a $n\times n$  unimodular polynomial matrix
so that its inverse $U^{-1}(z)$ is polynomial.\\
2) $G(z)$ is a $n\times p$ rational matrix having the form $G(z)=\bmat D(z)\\0\emat$ where $D(z)$ is a $p\times p$  diagonal matrix whose diagonal elements are non-zero rational functions  having only zeros strictly inside the unit circle.\\
3) $V(z)$ is a square unimodular polynomial matrix with $p$ rows and $p$ columns so that its inverse $V^{-1}(z)$ is also polynomial. Thus if $W(z)$ is minimum phase  the left inverse
$$W^{-L}(z) := V^{-1}(z)\  [D^{-1}(z)\mid 0] \ U^{-1}(z)$$
is clearly analytic.
Since there is an algorithm to compute the Smith-McMillan form, $W^{-L}(z)$ as defined above can be effectively computed.

\end{document}